%% file: paper2.tex
\documentclass[12pt]{article}

\usepackage{amsmath}
\usepackage{amssymb}
\usepackage{mathrsfs}
\usepackage[dvips]{graphicx}
\usepackage{color}

\newcommand{\beq}{\begin{equation}}
\newcommand{\bea}{\begin{eqnarray}}
\newcommand{\eeq}{\end{equation}}
\newcommand{\eea}{\end{eqnarray}}

\newcommand{\bx}{{\mathbf x}}

\begin{document}

\title{\bf Topological defect formation from 2PI effective action techniques}

\author{
J{\"u}rgen Berges, Stefan Roth\\[0.5cm]
Institute for Nuclear Physics\\
Darmstadt University of Technology\\
Schlossgartenstr. 9, 64289 Darmstadt, Germany}

\date{}
\begin{titlepage}
\maketitle
\def\thepage{}          

\begin{abstract}
We propose a quantum approach to nonequilibrium dynamics which combines the successful aspects of classical-statistical simulations on a lattice with the ability to take into account quantum corrections. It is based on the 2PI effective action for inhomogeneous fields and a volume average. This procedure does not involve any double counting which could appear in sampling prescriptions for inhomogeneous quantum evolutions. As an example, we study nonequilibrium dynamics of defect formation in $1+1$ dimensional relativistic scalar field theory and compare to insufficient descriptions based on homogeneous quantum fields. The latter are analyzed in detail by coupling the field to an external source, such that the emerging influence of defects can be studied by lowering the source to zero.
\end{abstract}

\end{titlepage}

\renewcommand{\thepage}{\arabic{page}}

\section{Introduction and outline}

Defect formation in symmetry breaking phase transitions is a nonequilibrium process with important applications for condensed matter physics or even early universe dynamics \cite{Topological}. It involves nonperturbative dynamics far from equilibrium which is a challenge for theoretical descriptions of quantum many body systems. Often the physics is approximated by classical-statistical field theory, which can be simulated on a lattice using Monte Carlo methods and numerical integration \cite{Khlebnikov:1996mc}. Classical Rayleigh-Jeans divergences and the lack of genuine quantum effects, such as the possibility to approach thermal equilibrium characterized by Bose-Einstein or Fermi-Dirac distributions, limit their use \cite{Berges:2004yj}. Lattice simulation methods for nonequilibrium quantum field theories, such as based on stochastic quantization \cite{Berges:2005yt}, are still in its infancies and it would be very desirable to be able to compute the impact of quantum corrections on defect formation using alternative approaches.

There has been substantial progress in recent years in the quantitative understanding of the nonequilibrium time evolution of quantum fields using approximate functional integral techniques. All information about a given nonequilibrium quantum field theory can be encoded in an effective action, which is the generating functional for all correlation functions. While naive perturbative expansions lead to secular behavior and fail to describe the time evolution, suitable approximations may be efficiently based on the two-particle irreducible (2PI) or higher $n$PI effective actions \cite{Berges:2004yj}.     
The nonperturbative 2PI $1/N$ expansion to next-to-leading order (NLO) \cite{Berges:2001fi}, where $N$ denotes the number of field components, has been particularly fruitful. 
This approach was applied to a variety of problems, related
to thermalization~\cite{Berges:2001fi,Cooper:2002qd,Tranberg:2008ae}, inflationary preheating \cite{Berges:2002cz}, fermion dynamics \cite{Berges:2002wr}, transport coefficients \cite{Aarts:2003bk}, nonthermal fixed points \cite{Berges:2008wm} and critical exponents of phase transitions \cite{Alford:2004jj}, or cold atoms \cite{Berges:2007ym}. The 2PI $1/N$ expansion has also been implemented in classical-statistical field theories, where it can be tested against full simulations on a lattice. Remarkable agreement was found between simulation results and those from the NLO approximation in various dimensions for not too small $N$ \cite{Aarts:2001yn}. Also NNLO corrections have been addressed~\cite{Aarts:2006cv}.

In contrast to those successful applications, it was pointed out in Ref.~\cite{Rajantie:2006gy} that the 2PI $1/N$ expansion to NLO does not reproduce known results of phase transition dynamics for $N=1$ and $N=2$. Of course, a $1/N$ expansion is not expected to work well for too small $N$ and quantitative deviations were observed before in this case \cite{Aarts:2001yn}. However, the results of Ref.~\cite{Rajantie:2006gy} suggest that the discrepancies arise because topological defects are present in this case, which are not properly described. 
This suggestion follows a long-standing discussion about the question of whether approximate functional integral techniques can describe topological defects. 

In this work we follow the nonequilibrium time evolution of the order parameter $\phi(t)$ and two-point correlation functions in a scalar $O(N)$-symmetric field theory with quartic self-interaction. This extends the study of Ref.~\cite{Rajantie:2006gy}, where defect formation from the dynamics of two-point correlation functions with $\phi \equiv 0$ is considered.  We further couple the field to an external source $J$. For large enough $J$ topological defects are suppressed
and we can study their emerging influence in detail by lowering $J$ for given $N$. We find that for not too small source there are only relatively small deviations between classical-statistical simulation results and those obtained from the 2PI $1/N$ expansion to NLO even for small $N$. On the other hand, as $J$ becomes sufficiently small such that topological defects become relevant, we find significant deviations. Our results confirm the conclusion that contributions from topological defects present for $N = 1$ and $N = 2$ are not properly described at NLO in the 2PI $1/N$ expansion. In general, we find that NLO results for $N=2$ agree somewhat better to those obtained from classical-statistical simulations than for $N=1$ as expected for a large-$N$ expansion. It would be very interesting to extend the study to NNLO though the computational effort becomes considerable~\cite{Aarts:2006cv} and alternative approaches can be more appropriate. 

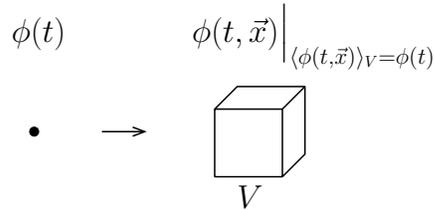
\begin{figure}
\begin{center}
\scalebox{0.6}{\input{average_phi.tex}}
\caption{The dynamics of the spatially homogeneous field expectation value $\phi(t)$ in the presence of defects can be replaced by a description using an inhomogeneous field $\phi(t,\bx)$
and a volume average. This allows one to study topological defect formation in quantum theories using inhomogeneous 2PI $1/N$ or loop expansions.}
\label{fig:fieldV}
\end{center}
\end{figure}
Here we propose an alternative quantum treatment, which allows one to quantitatively describe topological defect formation already at low orders in a $1/N$ expansion. At leading order the approach is similar to the inhomogeneous Hartree approximation considered in Ref.~\cite{Salle:2002fu}. It constructs the spatially homogeneous quantum system from a corresponding theory for an inhomogeneous "foreground" field $\phi(t,\bx)$ and a volume average as indicated in Fig.~\ref{fig:fieldV}. The additional spatial ${\bx}$-dependence of the field $\phi(t,\bx)$ is constrained by the fact that its volume average corresponds to the spatially homogeneous field expectation value of the original quantum theory:
\begin{equation}
\phi(t) = \langle \phi(t,\bx)\rangle_V \, ,
\label{eq:phif}
\end{equation}
where $\langle \ldots \rangle_V$ denotes an average over some local volume $V$ whose size is determined by the statistical error one wants to achieve.
Similarly, the higher $n$-point functions of the inhomogeneous description are related to those of the homogeneous quantum system by a volume average. We emphasize that without approximations the volume averaged inhomogeneous description agrees to the original system. In particular, this procedure does not involve any double counting which could appear in sampling prescriptions for inhomogeneous quantum evolutions. Here the description only exploits the fact that a spatially homogeneous system is by definition translation invariant in space.  We show that the classical-statistical approximation is included already in the inhomogeneous quantum approach to leading order in the 2PI $1/N$ expansion, while quantum corrections enter at next-to-leading order. 
 
The outline of the paper is as follows. In Sec.~\ref{sec:model} we describe classical kink and texture solutions in one spatial dimension for an $O(N)$ symmetric real scalar field theory. In Sec.~\ref{sec:classical_statistical_simulation} we include classical-statistical fluctuations by numerically integrating the classical field equations and Monte Carlo sampling. Sec.~\ref{sec:2PI_effective_action} introduces the quantum field theoretical description using the 2PI effective action. We discuss the role of quantum corrections and describe the classical-statistical limit. We consider in Sec.~\ref{sec:comparison_of_2PI_and_classical_statistical_simulations} the conventional approach, where the 2PI effective action is evaluated for a time-dependent but spatially homogeneous field expectation value $\phi(t)$. The results are compared with those from classical-statistical simulations for $N=1$ and $N=2$ for different $J$. In Sec.~\ref{sec:inhomogeneous_2PI} we introduce our inhomogeneous approach and compare with classical-statistical simulations.  Finally, in Sec.~\ref{sec:conclusion} we conclude and give an outlook.

\section{Topological defects in the $O(N)$ model}\label{sec:model}
 
We consider an $O(N)$ symmetric $N$-component real scalar field theory with quartic interaction in 1+1 dimensions following Ref.~\cite{Rajantie:2006gy}. In addition, we couple the classical fields $\varphi_a(t,x)$ ($a=1,\ldots,N$ with $t$ denoting time and $x$ is the variable for the spatial dimension) to explicitly symmetry breaking source terms $J_a$. The classical field equation of motion reads
\begin{equation}\label{eq:equation_of_motion}
 \partial_t^2\varphi_a +\gamma\partial_t\varphi_a -\partial_x^2\varphi_a +\mu^2\varphi_a +\frac{\lambda}{6N}(\varphi_b\varphi_b)\varphi_a - J_a = 0\, ,
\end{equation}
with mass parameter $\mu^2$ and quartic coupling $\lambda$. Summation over repeated indices is implied. The field equation of motion (\ref{eq:equation_of_motion}) contains a damping term $\gamma\partial_t\varphi_a(t,x)$ whose relevance will be discussed below.   

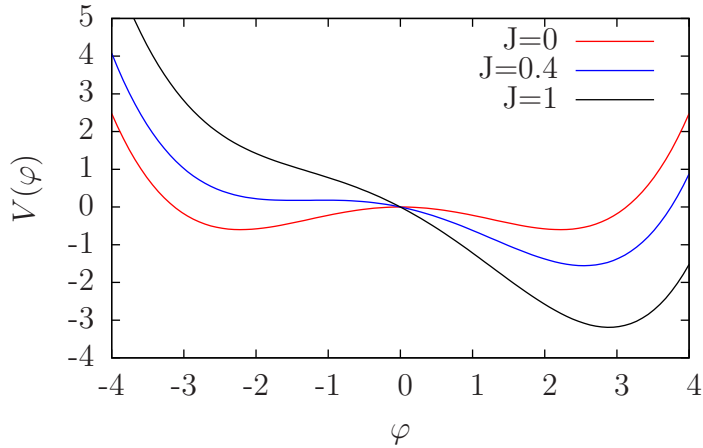
\begin{figure}[t]
 \begin{center}
  \input{field_analytic_potential}
  \caption[Classical potential of the scalar field]{Classical potential as a function of the field $\varphi$ for different values of the source $J$ for $N=1$. For later comparison with simulations in Sec.~\ref{sec:classical_statistical_simulation}, in lattice units of $a_s$ we use $\mu^2=-0.49$ and $\lambda=0.6$. Rising $J$ from zero lifts the degeneracy until only one minimum remains for $J > J^* \approx 0.42$.}
  \label{fig:analytic_potential}
 \end{center}
\end{figure}

The non-derivative part of (\ref{eq:equation_of_motion}) can be obtained from a quartic classical potential with linear source term,
\begin{equation}\label{eq:classicalV}
 V(\varphi) = \frac{1}{2}\mu^2\varphi_a\varphi_a+\frac{\lambda}{4!N}(\varphi_a\varphi_a)^2 - J_a \varphi_a \, .
\end{equation}
For the study of topological defect formation the initial classical potential is taken to be of free field form, 
\begin{equation}\label{eq:initial_potential}
 V_{\rm{initial}}(\varphi) =  \frac{1}{2}m^2\varphi_a\varphi_a,
\end{equation}
where $m^2>0$. At time $t=0^+$ the potential changes instantaneously to (\ref{eq:classicalV})
with $\mu^2=-m^2<0$, so the system is "quenched" into the phase with spontaneous symmetry breaking. For $N=1$ the classical potential then exhibits two degenerate minima at $\varphi = \pm v$ with
\begin{equation}\label{eq:v}
v =  |\mu| \sqrt{\frac{6}{\lambda}} 
\end{equation}
in the phase with spontaneous symmetry breaking for $J=0$. This is shown graphically in Fig.~\ref{fig:analytic_potential}.
The topological defect or kink solution interpolates between these separated minima. The stationary classical solution of
(\ref{eq:equation_of_motion}) reads
\begin{equation}\label{eq:kink_field}
 \varphi_{\rm{kink}}(x)= v \tanh\frac{x}{d} \, ,
\end{equation}
where $d=\sqrt{2}/|\mu|$ is the kink thickness. 
Below, we will study the behavior of kink solutions also for non-vanishing source $J$. From Fig.~\ref{fig:analytic_potential} one observes that by sufficiently increasing $J$ the potential starts to exhibit only one minimum. This happens around the critical value $J = J^*$ with
\begin{equation}\label{eq:Jstar}
J^* = \frac{2\, |\mu|^3}{3}\, \sqrt{\frac{2}{\lambda}} \, .
\end{equation}
Since kink solutions interpolate between two minima of the potential, they will be absent for $J \gtrsim J^*$, which is discussed in more detail in Sec.~\ref{sec:classical_statistical_simulation} on classical-statistical simulations.

For $N=2$ the minima of the classical potential in the phase with spontaneous symmetry breaking lie on a circle in the absence of sources. The classical equations of motion have texture solutions, which wind around the vacuum manifold and become indistinguishable from the vacuum in the infinite volume limit. Writing the two real field components as the sum of a real and imaginary part, texture solutions are
\begin{equation}
\varphi_R(x) + i \varphi_I(x) = v\, e^{2 \pi i N_w x/L} \, ,
\end{equation}
where the winding number $N_w$ is an integer. One observes the dependence of the texture solutions on the volume $L$. For $N > 2$ there are no topological defects for our model.

Finally, we will be interested in observables which are computed from ensemble averages.
The classical-statistical ensemble average is obtained from a normalized initial distribution $P[\varphi^0,\pi^0]$ with $\varphi^0_a(x) \equiv \varphi_a(t=0,x)$ and $\pi^0_a(x) = \partial_t \varphi_a(t,x)|_{t=0}$ such that
\begin{equation}
\langle \varphi_a(t,x) \rangle_{\rm{cl}} \, = \, \int {\cal D}\varphi^0 {\cal D}\pi^0 P[\varphi^0,\pi^0]\, \varphi_a(t,x)
\end{equation}
for the one-point function and accordingly for any $n$-point function. The measure indicates integration over classical phase space, 
$\int {\cal D}\varphi^0 {\cal D}\pi^0 = \int \prod_a \prod_x d\varphi^0_a(x) d\pi^0_a(x)$.
We consider spatially homogeneous ensembles relevant for early universe inflaton dynamics, such that the field average 
\begin{equation}\label{eq:macrofieldcl}
\phi_{a,{\rm{cl}}}(t) = \langle \varphi_a(t,x) \rangle_{\rm{cl}}
\end{equation} 
depends on time only. Similarly, the connected two-point function
\begin{equation}\label{eq:Fcl}
F_{ab,{\rm cl}}(t,t';x-y) = \langle \varphi_a(t,x) \varphi_b(t',y) \rangle_{\rm{cl}} 
- \phi_{a,{\rm{cl}}}(t) \phi_{b,{\rm{cl}}}(t')
\end{equation}
depends only on the relative coordinate. In Sec.~\ref{sec:classical_statistical_simulation} we will construct these ensemble averages from many individual solutions of (\ref{eq:equation_of_motion}) for initial conditions, which are sampled from a Gaussian initial distribution. We also note that the damping term $\gamma\partial_t\varphi$ in the equation of motion (\ref{eq:equation_of_motion}) is needed in order to be able to observe a non-vanishing field average (\ref{eq:macrofieldcl})
at late times, because of the absence of spontaneous symmetry breaking in a 1+1 dimensional system above zero temperature.

Since we will be interested in correlation functions, it is important to know how topological effects manifest themselves in correlators. A qualitative understanding may be obtained from considering the special case of randomly distributed kinks, which for $N=1$ may be represented in Fourier space by~\cite{Rajantie:2006gy}
\begin{equation}\label{eq:kink_propagator}
 F_{\rm{kink}}(p)= v^2 \, \frac{4 n}{4n^2+p^2}\left(\frac{\frac{1}{2}\pi p d}{\sinh(\frac{1}{2}\pi p d)}\right)^2.
\end{equation}
Here $n$ denotes the number density of kinks, which will be used as a fit parameter in analyzing numerical results from actual classical-statistical simulations. Correspondingly, 
for $N=2$ we employ~\cite{Rajantie:2006gy}  
\begin{equation}\label{eq:texture_propagator}
 F_{\rm{texture}}(p)= v^2\sqrt{\frac{6\xi L}{\pi}}\exp\left(-\frac{3\xi L}{2\pi^2}p^2\right) \, ,
\end{equation}
where $v$ is taken to denote the same value as in (\ref{eq:v}) and the length scale $\xi$ will be taken as a fit parameter.

\section{Classical-statistical simulation}
\label{sec:classical_statistical_simulation}

In this section we consider the classical field theory on a spatial lattice. In the following, all quantities are expressed in appropriate units of the spatial lattice spacing $a_s$. If not stated otherwise we choose a lattice with length $L=256$.  For the time evolution we discretize the time in steps of typically $\Delta t=0.1$. The other relevant parameters are the mass parameter $\mu^2=-0.49$, coupling constant $\lambda=0.6$ and damping rate $\gamma=0.4$.

We sample the initial conditions for the classical equation of motion \eqref{eq:equation_of_motion}, where the initial fields $\varphi_a(t=0,p) = A_a(p) \exp(i \alpha_a(p))$ are taken in spatial Fourier space from a Gaussian ensemble for $A_a(p)$ with a random complex phase $\alpha_a(p)$ respecting $\varphi_a(t,p)=\varphi_a^*(t,-p)$, and similarly for $\partial_t\varphi_a(t,p)|_{t=0}$. As a consequence, the initial field average (\ref{eq:macrofieldcl}) and also its initial time derivative is taken to vanish: 
\begin{equation}
\phi_{a,{\rm{cl}}}(t=0) \, = \, 0 \quad , \quad \partial_t \phi_{a,{\rm{cl}}}(t)|_{t=0} \, = 0 \, .
\label{eq:fieldinitialcl}
\end{equation}
The initial two-point function (\ref{eq:Fcl}) and derivatives are of free-field form. In Fourier space they read
\begin{eqnarray} \label{eq:field_inital_conditions}
   F_{ab,{\rm{cl}}}(t,t';p)|_{t=t'=0} & = & \frac{1}{2\omega_{p}}\, \delta_{ab} 
   \, ,\nonumber \\
    \partial_{t}F_{ab,{\rm{cl}}}(t,t';p)|_{t=t'=0} & = & 0 \, ,\\
    \partial_{t}\partial_{t'}F_{ab,{\rm{cl}}}(t,t';p)|_{t=t'=0} & = & \frac{\omega_{p}}{2} \, \delta_{ab} \, ,\nonumber  
\end{eqnarray}
where $\omega_p=\sqrt{p^2+m^2}$ with $m^2 = -\mu^2$. We observe good convergence by sampling the initial conditions over 1000 runs and exploit spatial translation invariance to further improve statistics by taking lattice averages.  

\begin{figure}[t]
 \begin{center}
  \input{symmetry_breaking}
  \caption[Classical field average]{The classical field average
  $\phi_{\rm{cl}}$ as a function of the source $J$ at late times for $N=1$. As $J$ goes to zero a non-vanishing $\phi_{\rm{cl}}$ is observed for the damped case ($\gamma = 0.4$), while $\phi_{\rm{cl}}$ approaches zero in the absence of damping ($\gamma = 0$) as expected. For comparison, the solid curve shows the solution obtained from minimizing the classical potential (\ref{eq:classicalV}).}
  \label{fig:averaged field}
 \end{center}
\end{figure}
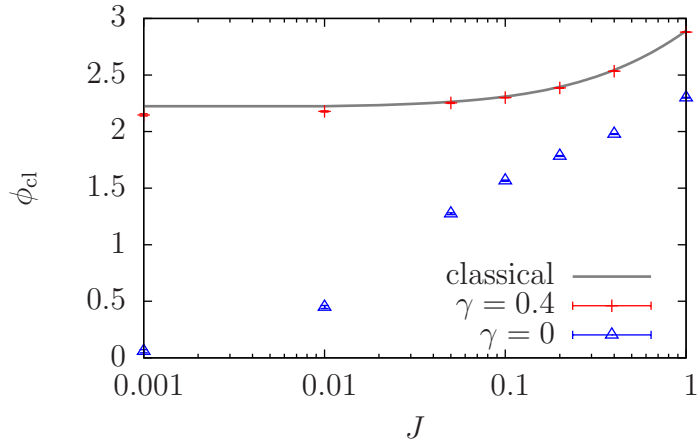 
We first compare the classical-statistical evolution for the undamped ($\gamma = 0$) and damped ($\gamma > 0$) case for $N=1$. Fig.~\ref{fig:averaged field} shows the classical field average $\phi_{\rm{cl}}$ given by (\ref{eq:macrofieldcl}) as a function of the homogeneous source $J$ at late times. As expected, in the absence of damping no spontaneous symmetry breaking with $\phi_{\rm{cl}} \neq 0$ is observed in the limit $J \to 0$. In contrast, the result with damping ($\gamma = 0.4$) exhibits spontaneous symmetry breaking and defect formation may be observed in this case. We find that for the considered parameters the field expectation value is rather accurately described by the value of the minimum of the classical potential $v \simeq 2.21$ given by (\ref{eq:v}).

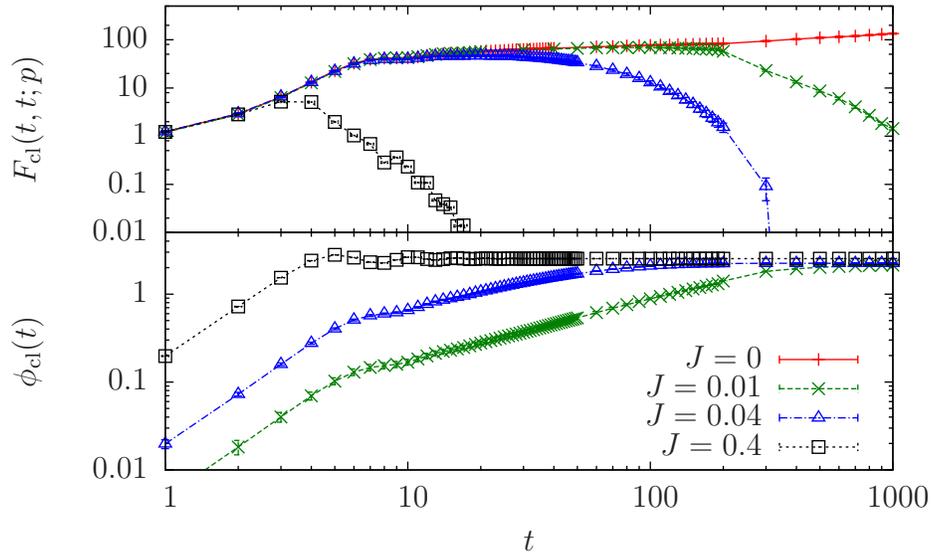
\begin{figure}[t]
 \begin{center}
  \input{CLSTAT_phi_F}
  \caption[First impulse-mode for different sources]{Upper graph: The equal-time two-point correlator as a function of time, spatially Fourier transformed and for different sources $J$ with $N=1$, $\gamma = 0.4$. Lower graph: The classical field average as a function of time. For identically vanishing source ($J=0$), $\phi_{\rm{cl}}$ remains zero by symmetry.}
  \label{fig:wo_source_g04_p1}
 \end{center}
\end{figure}
We are particularly interested in the classical-statistical correlation function (\ref{eq:Fcl})
at equal time, $t=t'$. According to (\ref{eq:kink_propagator}), a stationary value for this quantity in Fourier space should give direct information about the kink density. The upper graph of Fig.~\ref{fig:wo_source_g04_p1} shows for $N=1$ the spatially Fourier transformed correlation function for low momentum, $p \simeq 0$, as a function of time for different values of the source $J$ and $\gamma = 0.4$. After an initial fast growth, one observes a comparably slow evolution for small $J$. With increasing source this quasi-stationary period is diminished and clearly absent for $J \gtrsim 0.4$. The latter value is close to $J^*$ given by (\ref{eq:Jstar}) for which the classical potential no longer exhibits two distinct minima (see Fig.~\ref{fig:analytic_potential}) and defect formation is not expected in this case. 

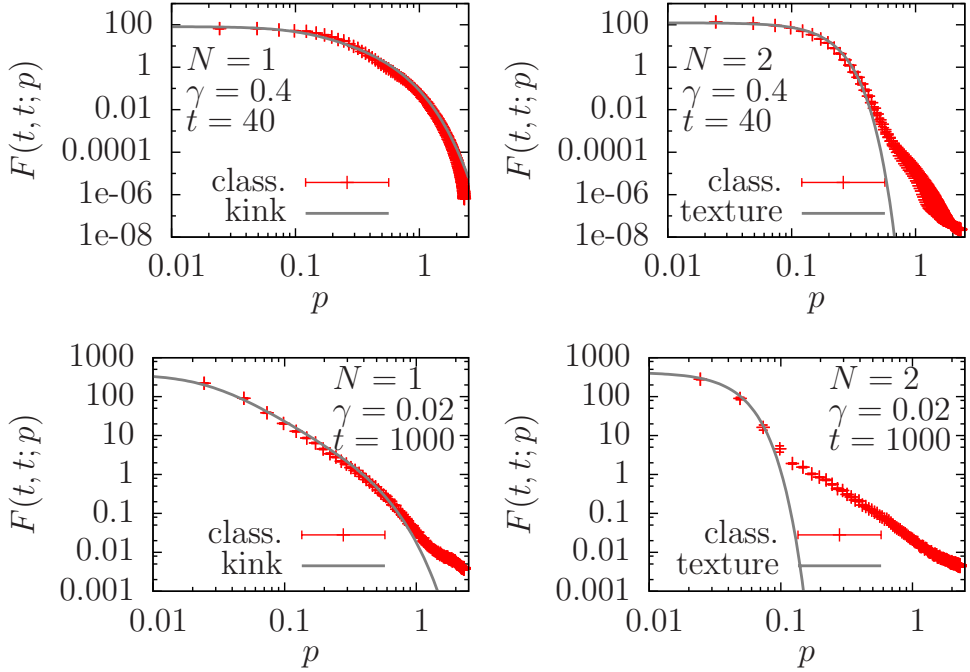
\begin{figure}[t]
 \begin{center}
\scalebox{1.06}{\input{defects}}
  \caption[Defect solutions]{Classical simulation results for the equal-time two-point function and fits to the corresponding defect ans{\"a}tze (solid lines) are shown as a function of momentum. Upper graphs are for $\gamma=0.4$ and lower graphs for $\gamma=0.02$ both for $N=1$ (left) and $N=2$ (right). The times are chosen to be within the quasi-stationary period of the two-point function. Data and fits agree well for lower momenta, indicating the presence of topological defects.}
  \label{fig:defects}
 \end{center}
\end{figure}
The lower graph of Fig.~\ref{fig:wo_source_g04_p1} shows $\phi_{\rm{cl}}$ as a function of time for different values of $J$. The field dynamics for $J = 0.01$ is already found to approximate well the limit $J \to 0^+$, while for $J \equiv 0$ the classical field average vanishes as it should by symmetry. Similar to what is observed for the two-point function, $\phi_{\rm{cl}}$ exhibits three characteristic time regimes. After an initial power-law growth, the dynamics slows down considerably. Finally, $\phi_{\rm{cl}}$ saturates to the order-parameter value in the spontaneously broken phase (see Fig.~\ref{fig:averaged field}).
The field average takes on its maximum value when all configurations belong to the same phase and the phase transition is completed. No kink contributions, which interpolate between different field values, are expected in this case. Accordingly, by comparison with the upper graph one observes that the time when $\phi_{\rm{cl}}$ reaches its maximum is also the time when the equal-time $F_{\rm{cl}}$ exits the quasi-stationary period and starts to drop rapidly. A very similar behavior can also be observed for $N = 2$.    

It remains to show that the intermediate quasi-stationary period for the two-point function is indeed characterized by topological defects. Fig.~\ref{fig:defects} shows the correlator as a function of momentum for fixed time and $J = 0$. The upper graphs are for $\gamma = 0.4$ and $t= 40$. At this time the evolution of the two-point function is in the quasi-stationary regime both for $N=1$ (left) and $N=2$ (right). The fits using the kink ansatz (\ref{eq:kink_propagator}) with $n = 0.06$ and (\ref{eq:texture_propagator}) with $\xi = 1.3$ for textures are also shown. Simulation results and analytical expressions agree well in the infrared, clearly indicating the presence of topological defects. For comparison, we give in addition results for smaller damping, $\gamma = 0.02$. In this case the evolution for the correlator enters the quasi-stationary regime for the equal-time two-point function at a later time beginning around $t=100$. The lower graphs show the results at $t=1000$, which again indicate the presence of defects. The classical-statistical simulation results will be compared with those from the 2PI effective action below, where we will also further discuss the $J$-dependence.

\section{Quantum theory: 2PI $1/N$ to NLO}
\label{sec:2PI_effective_action}

To compute the quantum equations of motion for our damped model of Sec.~\ref{sec:model}, it is useful to note that the classical field equation (\ref{eq:equation_of_motion}) can also be viewed as the equation of motion of an undamped 2+1 dimensional scalar field theory in an expanding anisotropic geometry with the expansion factor $a(t)=a_0\exp(\gamma t)$ such that $\gamma \equiv \dot{a}/a$ \cite{Rajantie:2006gy}. This allows one to use standard 2PI effective action techniques for the quantum theory, which do not suffer from secularity problems and where nonperturbative approximations are available such as the 2PI large-$N$ expansion to NLO~\cite{Berges:2001fi}. Further details can also be found in appendix \ref{app:Calculation of the 2PI action}. 

We consider the $O(N)$-symmetric quantum field theory with quartic self-interaction corresponding to the classical model of Sec.~\ref{sec:model}. For Heisenberg field operators $\hat{\varphi}_a(t,x)$, we compute the quantum equations of motion for one- and two-point functions
\begin{eqnarray}
 \phi_a(t,x) & = & \langle\hat{\varphi}_a(t,x)\rangle \, ,\\
 \rho_{ab}(t,t';x,y) & = & i\langle\left[\hat{\varphi}_a(t,x),\hat{\varphi}_b(t',y)\right]\rangle \, ,\\
 F_{ab}(t,t';x,y) & = & \frac{1}{2}\langle\left\{\hat{\varphi}_a(t,x),\hat{\varphi}_b(t',y)\right\}\rangle
 - \phi_a(t,x) \phi_b(t',y) \, .
\end{eqnarray}
Here $\phi$ denotes the field expectation value, $\rho$ is the spectral function determined by the commutator and $F$ is the statistical propagator obtained from the anti-commutator of two field operators. Even though we are finally interested in spatially homogeneous systems, we consider for a moment the equations of motion for inhomogeneous fields and correlation functions to be used in Sec.~\ref{sec:inhomogeneous_2PI}. 

As detailed in appendix \ref{app:Calculation of the 2PI action}, at NLO in the 2PI $1/N$ expansion one obtains the field evolution equation~\cite{Berges:2001fi}:
\begin{eqnarray}  && \left[\left(\partial_t^2+\gamma\partial_t-\partial_x^2+\frac{\lambda}{6N}\phi^2(t,x)\right) \delta_{ab} + M^2_{ab}(t,x;\phi = 0) \right]\phi_b(t,x) \nonumber\\
&&  = \, J_a(t,x) - \int_{0}^{t}dt'\int dy\;a(t')\Sigma^{\rho}_{ab}(t,t';x,y;\phi=0)\phi_b(t',y) \, , 
\label{eq:quantumphi}
\end{eqnarray}
where the initial time is taken as $t=0$ and $\phi^2 \equiv \phi_a \phi_a$.
Compared to the classical equation of motion \eqref{eq:equation_of_motion}, the effect of fluctuations in the quantum description enter in terms of the mass-like term $M^2$ and self-energy $\Sigma^{\rho}$ which are given below. The corresponding evolution equations for $F$ and $\rho$ are~\cite{Berges:2001fi}
\begin{eqnarray} \lefteqn{\left[\left(\partial_t^2+\gamma\partial_t-\partial_x^2\right)\delta_{ab}+M_{ab}^2(t,x)\right]F_{bc}(t,t';x,y)}\notag\\
    & = & -\int_{0}^{t}dt''\int dz\;a(t'')\Sigma^{\rho}_{ab}(t,t'';x,z)F_{bc}(t'',t';z,y)\notag\\
    & & +\int_{0}^{t'}dt''\int dz\;a(t'')\Sigma^F_{ab}(t,t'';x,z)\rho_{bc}(t'',t';z,y)\, ,\notag\\ \lefteqn{\left[\left(\partial_t^2+\gamma\partial_t-\partial_x^2\right)\delta_{ab}+M_{ab}^2(t,x)\right]\rho_{bc}(t,t';x,y)}\notag\\
    & = & -\int_{t'}^{t}dt''\int dz\;a(t'')\Sigma^{\rho}_{ab}(t,t'';x,z)\rho_{bc}(t'',t';z,y) \, .\label{eq:Frho}
\end{eqnarray}
At NLO in the 2PI $1/N$ expansion the mass and self-energy terms appearing in (\ref{eq:quantumphi}) and (\ref{eq:Frho}) are~\cite{Berges:2001fi}
\begin{eqnarray}\label{eq:mass_squared}
&& M_{ab}^2(t,x) \, = \, \left(\mu^2+\frac{\lambda}{6N}\left[F_{cc}(t,t;x,x) + \phi^2(t,x) \right] \right)\delta_{ab} \nonumber\\ 
&& +\frac{\lambda}{3N} \left[ F_{ab}(t,t;x,x) + \phi_a(t,x)\phi_b(t,x) \right]\, , \nonumber \\ \label{eq:sigma_f}
&& \Sigma^F_{ab}(t,t';x,y) \, = \,  -\frac{\lambda}{3N}\bigg(I_F(t,t';x,y)\left[\phi_a(t,x)\phi_b(t',y)+F_{ab}(t,t';x,y)\right]  \notag\\
&& -\frac{1}{4}I_{\rho}(t,t';x,y)\rho_{ab}(t,t';x,y) +P_F(t,t';x,y)F_{ab}(t,t';x,y) \notag\\
&& -\frac{1}{4}P_{\rho}(t,t';x,y)\rho_{ab}(t,t';x,y)\bigg) \, , \nonumber\\
&& \Sigma^{\rho}_{ab}(t,t';x,y) \, = \,  -\frac{\lambda}{3N}\bigg(I_{\rho}(t,t';x,y)\left[\phi_a(t,x)\phi_b(t',y)+F_{ab}(t,t';x,y)\right] \notag\\
&& +I_F(t,t';x,y)\rho_{ab}(t,t';x,y)+P_{\rho}(t,t';x,y)F_{ab}(t,t';x,y) \notag\\
&& +P_F(t,t';x,y)\rho_{ab}(t,t';x,y)\bigg)  \, ,
\end{eqnarray}
with the summation functions
\begin{eqnarray}\label{eq:i_f}
&& I_F(t,t';x,y) \, = \, \frac{\lambda}{6N}\Bigg(\left[F^2(t,t';x,y)-\frac{1}{4}\rho^2(t,t';x,y)\right]\notag\\
    & & -\int_{0}^{t}dt''\int dz\;a(t'')I_{\rho}(t,t'';x,z)\left[F^2(t'',t';z,y)-\frac{1}{4}\rho^2(t'',t';z,y)\right]\notag\\
    & & + 2\int_{0}^{t'}dt''\int dz\;a(t'')I_F(t,t'';x,z)F_{ab}(t'',t';z,y)\rho_{ab}(t'',t';z,y)\Bigg)\, ,\notag\\
&& I_{\rho}(t,t';x,y) \, = \, \frac{\lambda}{3N}F_{ab}(t,t';x,y)\rho_{ab}(t,t';x,y)\notag\\
    & & -\frac{\lambda}{3N}\int_{t'}^{t}dt''\int dz\;a(t'')I_{\rho}(t,t'';x,z)F_{ab}(t'',t';z,y)\rho_{ab}(t'',t';z,y)\notag \, ,
\end{eqnarray}
where $F^2 \equiv F_{ab} F_{ab}$, etc. For non-vanishing field one also has
\begin{eqnarray}\label{eq:P}
    && \!\!\!\! P_F(t,t';x,y) =  -\frac{\lambda}{3N}\Bigg(H_F(t,t';x,y) -\int_{0}^{t}dt''\int dz\,a(t'') [H_{\rho}(t,t'';x,z) \notag\\    &&\times I_F(t'',t';z,y)+I_{\rho}(t,t'';x,z)H_F(t'',t';z,y)] +\int_{0}^{t'}dt''\int dz\, a(t'') \notag\\
    & & \!\!\!\! \times  \left[H_F(t,t'';x,z)I_{\rho}(t'',t';z,y) + I_F(t,t'';x,z)H_{\rho}(t'',t';z,y)\right] \notag\\
    & & \!\!\!\! -\int_{0}^{t}\!\! dt'' \! \int_{0}^{t'} \!\! dt''' \! \int \! dz d\nu\, a(t'') a(t''') I_{\rho}(t,t'';x,z) H_F(t'',t''';z,\nu) I_{\rho}(t''',t';\nu,y) \notag\\
    & & \!\!\!\! +\int_{0}^{t}\!\! dt'' \! \int_{0}^{t''} \!\! dt''' \! \int \! dz d\nu\, a(t'') a(t''') I_{\rho}(t,t'';x,z) H_{\rho}(t'',t''';z,\nu) I_F(t''',t';\nu,y) \notag\\
    & & \!\!\!\! +\int_{0}^{t'}\!\! dt'' \! \int_{t''}^{t'} \!\! dt''' \! \int \! dz d\nu a(t'') a(t''') I_F(t,t'';x,z) H_{\rho}(t'',t''';z,\nu) I_{\rho}(t''',t';\nu,y)
 \Bigg) , \notag\\
    && \!\!\!\! P_{\rho}(t,t';x,y) = -\frac{\lambda}{3N}\Bigg(H_{\rho}(t,t';x,y) -\int_{t'}^{t}dt''\int dz\;a(t'')\notag\\
    & & \!\!\!\! \times\left[H_{\rho}(t,t'';x,z)I_{\rho}(t'',t';z,y)+I_{\rho}(t,t'';x,z)H_{\rho}(t'',t';z,y)\right]\notag\\
    & & \!\!\!\! +\int_{t'}^{t} \!\! dt'' \int_{t'}^{t''} \!\! dt''' \! \int \! dz d\nu \, a(t'')a(t''')I_{\rho}(t,t'';x,z)H_{\rho}(t'',t''';z,\nu) I_{\rho}(t''',t';\nu,y)\Bigg) \, ,\notag
\end{eqnarray}
with
\begin{eqnarray}\label{eq:H}
    H_F(t,t';x,y) & = & -\phi_a(t,x)F_{ab}(t,t';x,y)\phi_b(t',y) \, , \nonumber \\
    H_{\rho}(t,t';x,y) & = & -\phi_a(t,x)\rho_{ab}(t,t';x,y)\phi_b(t',y) \, .
\end{eqnarray}
We note that the terms $-\frac{1}{4}I_{\rho}\rho$ and $-\frac{1}{4}P_{\rho}\rho$ in \eqref{eq:sigma_f} and the terms proportional to $\rho^2$ in the summation function $I_F$ describe genuine quantum corrections, whereas all other terms would also be present in a classical-statistical description of the dynamics~\cite{Aarts:2001yn}. Neglecting these terms one obtains, therefore, the evolution equations for the corresponding classical-statistical field theory~\cite{Berges:2004yj}.

The above evolution equations have to be supplemented by initial conditions for the field $\phi(t,x)$ and its first derivatives, and similarly for the correlation functions. For the spectral function these are determined by the boson field commutation relations, i.e.
\begin{eqnarray}
    \rho_{ab}(t,t';x,y)|_{t=t'} & = & 0\, ,\nonumber\\
    \partial_t\rho_{ab}(t,t';x,y)|_{t=t'} & = & \frac{\delta(x-y)}{a(t)}\, \delta_{ab}\, ,
    \label{eq:initialrho}\\
    \partial_t\partial_{t'}\rho_{ab}(t,t';x,y)|_{t=t'} & = & 0 \, ,\nonumber
\end{eqnarray}
which are valid at all equal times including $t=t'=0$.
The initial conditions for the statistical propagator for homogeneous systems are chosen to be of the free-field form in accordance with (\ref{eq:field_inital_conditions}) of Sec.~\ref{sec:model}:
\begin{eqnarray}
    F_{ab}(t,t';p)|_{t=t'=0} & = & \frac{1}{2\omega_{p}}\, \delta_{ab} \, ,\nonumber \\
    \partial_{t}F_{ab}(t,t';p)|_{t=t'=0} & = & 0 \, , \label{eq:initialF}\\
    \partial_{t}\partial_{t'}F_{ab}(t,t';p)|_{t=t'=0} & = & \frac{\omega_{p}}{2} \, \delta_{ab} \, ,\nonumber 
\end{eqnarray}
where again $\omega_p=\sqrt{p^2+m^2}$ with $m^2 = -\mu^2$. The initial field average and also its initial time derivative are taken to vanish as in (\ref{eq:fieldinitialcl}), 
\begin{equation}
\phi_{a}(t=0) \, = \, 0 \quad , \quad \partial_t \phi_{a}(t)|_{t=0} \, = \, 0 \, .
\label{eq:fieldinitial}
\end{equation}
By considering $O(N)$-symmetric initial conditions as above, the evolution equations can be simplified. Rotating the field appropriately we are only left with one non-vanishing field component and the propagators reduce to one parallel and $(N-1)$ times the same perpendicular component.

For comparison of results from the quantum evolution equations and from the classical-statistical simulations of Sec.~\ref{sec:classical_statistical_simulation}, we regularize the theory on a spatial lattice with periodic boundary conditions and consider discretized time steps $\Delta t$. Typically $\Delta t=0.1$ is sufficient to observe insensitivity of results to time-discretization errors. We will chose the same parameters as in Sec.~\ref{sec:classical_statistical_simulation} if not stated otherwise.

\section{2PI $1/N$ expansion for homogeneous fields}
\label{sec:comparison_of_2PI_and_classical_statistical_simulations}

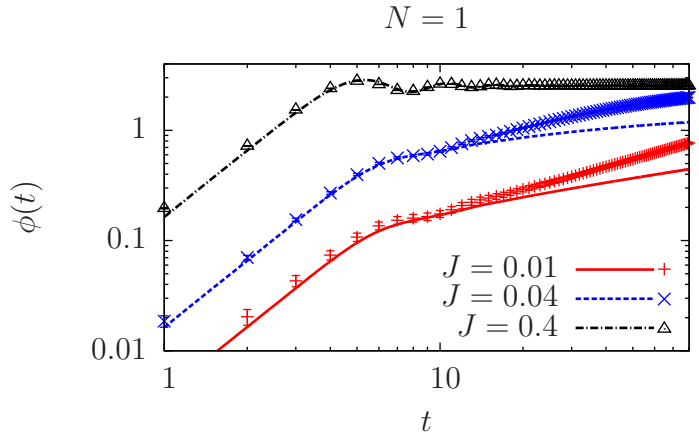
\begin{figure}
 \begin{center}
\input{new_field_final_N1_phi_log}
  \caption[Field expectation value for N=1]{Comparison of the evolution of the field from the 2PI $1/N$ expansion to NLO (lines) and classical-statistical simulations (symbols) for $N=1$ and different values of the source $J$.}
  \label{fig:new_field_final_N1_phi_log}
 \end{center}
\end{figure}
In the classical-statistical simulations of Sec.~\ref{sec:classical_statistical_simulation} we considered different individual realizations of the classical field. Each realization is inhomogeneous and after averaging over many members of a given ensemble the homogeneous field average is obtained. In contrast, the quantum description of Sec.~\ref{sec:2PI_effective_action} evolves the field expectation value using (\ref{eq:quantumphi}), which includes already the quantum fluctuations and requires no further averaging. Therefore, for spatially homogeneous quantum systems the time-dependent field $\phi(t)$ has no spatial dependence. It is pointed out in Ref.~\cite{Rajantie:2006gy} that in the presence of topological defects an approximate description based on a $1/N$ expansion can be inaccurate in this case. The results are based on a comparison of classical-statistical simulations as in Sec.~\ref{sec:classical_statistical_simulation} with solutions of the NLO evolution equations (\ref{eq:Frho}) in the classical-statistical field theory limit as described in Sec.~\ref{sec:2PI_effective_action}. In Ref.~\cite{Rajantie:2006gy} the field expectation value is taken to vanish. However, $\phi \equiv 0$ does not correspond to the true field value in a symmetry breaking phase transition. In particular, the evolution equations (\ref{eq:Frho}) for the correlation functions $F$ and $\rho$ receive corrections from a non-zero $\phi$.  In this section, we extend that study by following the time evolution of the non-zero order parameter $\phi(t)$ and for non-zero source $J$. After analyzing in some detail the shortcomings of the 2PI effective action approach for homogeneous fields to describe physics of topological defects, we proceed in Sec.~\ref{sec:inhomogeneous_2PI} with an inhomogeneous 2PI description resolving these difficulties.

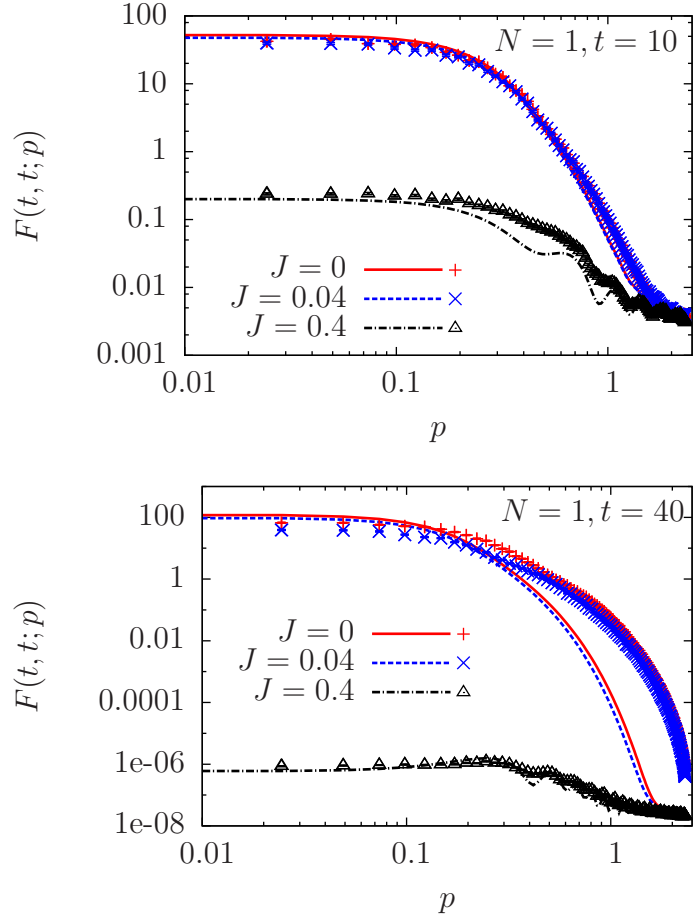
\begin{figure}
 \begin{center}
\input{new_field_final_N1_t10}
\input{new_field_final_N1_t40}
  \caption[Field expectation value for N=1]{The statistical two-point function as a function of momentum for fixed times $t=10$ (top) and $t=40$ (down) for $N=1$ and different values of the source $J$. Compared are NLO results (lines) and those from classical-statistical simulations (symbols).}
  \label{fig:new_field_final_N1}
 \end{center}
\end{figure}
We compute the nonequilibrium time evolution for the quantum system solving (\ref{eq:quantumphi})--(\ref{eq:H}) using homogeneity and isotropy, such that the field $\phi(t,x) = \phi(t)$ and the statistical correlator $F(t,t';x,y) = F(t,t';|x-y|)$, and equivalently for the spectral function $\rho$. The resulting equations are solved in spatial Fourier space, with the initial conditions $\phi(t=0)=\partial_t \phi(t)|_{t=0}=0$ as well as (\ref{eq:initialrho}) and (\ref{eq:initialF}) for the correlators. We have explicitly verified from the numerical solutions that classical-statistical fluctuations dominate over quantum fluctuations for all times that are relevant for topological defect formation.\footnote{Of course, genuine quantum corrections are crucial for the asymptotic late-time approach to thermal equilibrium, which has been studied elsewhere~\cite{Berges:2001fi,Cooper:2002qd,Berges:2002wr}.} In this case all $F^2$ terms dominate over $\rho^2$ terms in the above evolution equations. As mentioned in Sec.~\ref{sec:2PI_effective_action} this represents the classical-statistical limit of these equations~\cite{Aarts:2001yn}, which we employ in the following. As a consequence, the 2PI approach and the 'exact' classical-statistical simulations describe the very same theory and differences arise solely because of approximations. 

\begin{figure}
\begin{center}
\input{new_field_final_N2_phi_log}
\caption[Field expectation value for N=2]{Shown is the non-zero component of field expectation value as in Fig.~\ref{fig:new_field_final_N1_phi_log} but for $N=2$.}
\label{fig:new_field_final_N2_phi_log}
\end{center}
\end{figure}
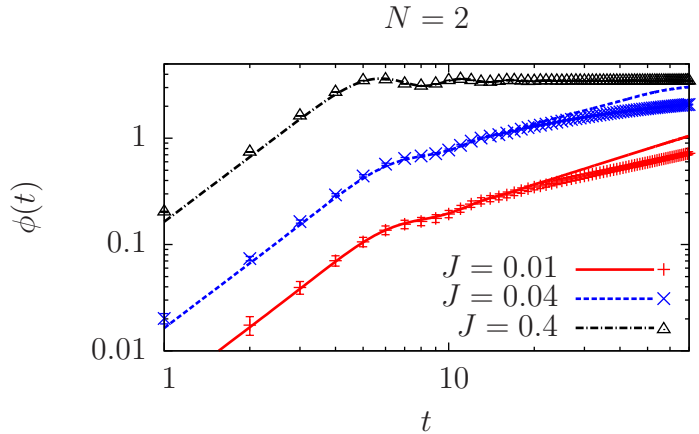
Fig.~\ref{fig:new_field_final_N1_phi_log} shows the time evolution of the field $\phi(t)$ for $N=1$ and various values of the source $J$. The results from the 2PI $1/N$ expansion to NLO are drawn as lines while those from the corresponding classical-statistical simulation are represented by symbols. According to the discussion of Sec.~\ref{sec:classical_statistical_simulation}, for our model we expect significant contributions from defects starting around $t\gtrsim 10$ after which the equal-time two-point function $F(t,t;p)$ enters a quasi-stationary regime. This quasi-stationary period for $F(t,t;p)$ is absent for sources exceeding $J^* \simeq 0.4$ and lasts until about $t=40$ for $J=0.04$ (see Fig.~\ref{fig:wo_source_g04_p1}). We plot in Fig.~\ref{fig:new_field_final_N1_phi_log} the field evolution until $t= 80$. One observes that the initial power-law growth until $t \simeq 10$ is accurately reproduced by the NLO approximation. A rather good agreement is also found at later times for $J = 0.4$, for which defects are expected to play no role. This agreement is remarkable by itself since there is no a priori reason that a $1/N$ expansion evaluated at $N=1$ should be accurate. In contrast, for smaller $J$ significant deviations between classical-statistical simulation and NLO results are building up quickly.  

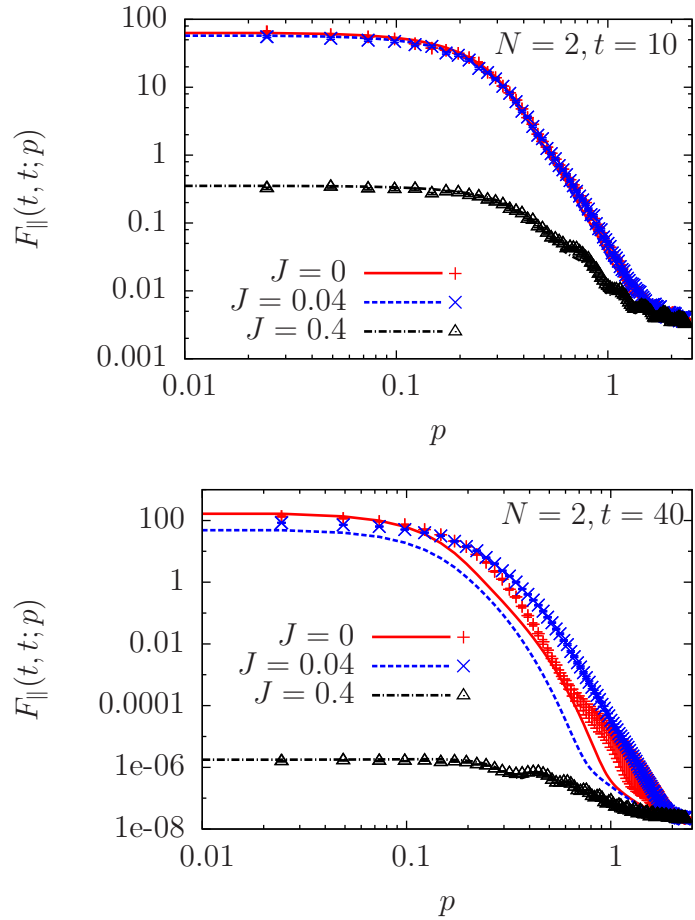
\begin{figure}
\begin{center}
\input{new_field_final_N2_t10}
\input{new_field_final_N2_t40}
\caption[Statistical propagator at t=10/40, N=2]{The same as in Fig.~\ref{fig:new_field_final_N1} for $N=2$. Shown is the longitudinal part of the statistical two-point function.}
\label{fig:new_field_final_N2_t}
\end{center}
\end{figure}
\begin{figure}
\begin{center}
\input{new_field_final_N2_t10_trans}
\input{new_field_final_N2_t40_trans}
\caption[Statistical propagator at t=10/40, N=2,trans]{The same as in Fig.~\ref{fig:new_field_final_N2_t} but for the transverse part of the statistical two-point function.}
\label{fig:new_field_final_N2_trans_t}
\end{center}
\end{figure}
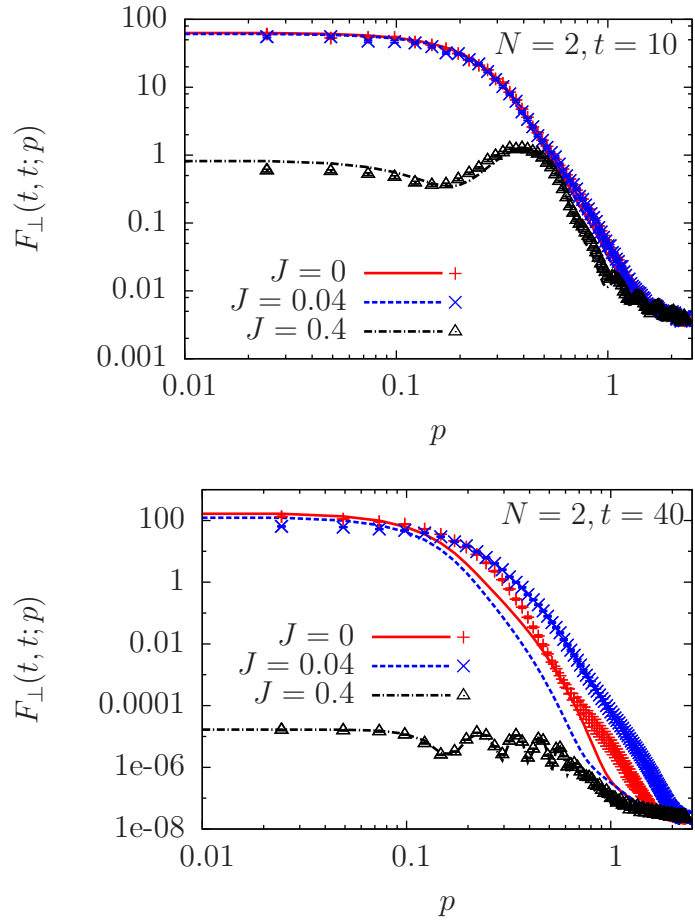
A similar finding can also be obtained from comparisons of the results for the statistical two-point function. Fig.~\ref{fig:new_field_final_N1} shows $F(t,t;p)$ as a function of spatial momentum for fixed times and various values of $J$ for $N=1$. The upper graph at $t=10$ confirms a rather good agreement between the NLO approximation and simulation results for small enough $J$. Only for $J = 0.4$ some deviations are observed at $t=10$ at which $F(t,t;p)$ started already to decay significantly according to Fig.~\ref{fig:wo_source_g04_p1} of Sec.~\ref{sec:classical_statistical_simulation}. In this region $F(t,t;p)$ oscillates rapidly, so a slight phase difference between the two computations can lead to a pronounced difference.
At later times this dynamics slows down and the observed deviations become smaller again. This can be seen from the lower graph of Fig.~\ref{fig:new_field_final_N1} showing results at $t=40$. In contrast, for smaller $J$ one observes increased deviations at $t= 40$ as compared to earlier times. Comparison with Fig.~\ref{fig:defects} of Sec.~\ref{sec:classical_statistical_simulation} shows that the NLO results do not reproduce well the shape of $F(t,t;p)$ which is characteristic for contributions from defects for small $J$.   

The same analysis is also done for $N=2$. The qualitatively similar results are shown in Figs.~\ref{fig:new_field_final_N2_phi_log} to \ref{fig:new_field_final_N2_trans_t}. In general, we find a somewhat better agreement between the NLO approximation and classical-statistical simulations as compared to $N=1$, which should be expected for a $1/N$ expansion. We have verified that for $N > 2$ the agreement becomes excellent, as is reported as well in Ref.~\cite{Rajantie:2006gy}. However, no defects are present in this case. In view of these results, it would be very interesting to perform a comparison with the 2PI $1/N$ expansion to NNLO. However, the computational effort becomes considerable at NNLO~\cite{Aarts:2006cv} and alternative approaches may be more appropriate, which is discussed in the following.

\section{2PI expansions with inhomogeneous fields}
\label{sec:inhomogeneous_2PI}

In Sec.~\ref{sec:comparison_of_2PI_and_classical_statistical_simulations} the 2PI effective action is considered for a time-dependent but spatially homogeneous field expectation value $\phi(t)$. This reflects the spatial homogeneity of the considered quantum system. Apparently, this is an inefficient starting point for a $1/N$ expansion if physics of inhomogeneous classical field configurations such as topological defects is involved. In classical-statistical simulations each realization of the field is inhomogeneous and only after averaging over many members of a given ensemble the homogeneous field is obtained. In contrast, the quantum 2PI effective action does not give information about individual realizations but only about expectation values or quantum-statistical averages. It is also not obvious that one could consider individual evolutions of inhomogeneous quantum field expectation values whose ensemble average would give the correct homogeneous field without "double counting". 

Here we propose an approach, which takes into account the fact that defect formation is most efficiently described using inhomogeneous fields but without the need of sampling techniques such that no double counting problem for quantum theories can occur. In this prescription the spatially homogeneous quantum system is constructed from a corresponding inhomogeneous system and a volume average. More precisely, the homogeneous field $\phi(t)$ is replaced by an inhomogeneous field $\phi(t,x)$ with an additional spatial $x$-dependence. Similarly, the spatially homogeneous two-point correlation functions such as $F(t,t';x-y)$ are represented by their inhomogeneous counterparts, $F(t,t';x,y)$, etc. The inhomogeneous field and correlation functions are governed by the quantum evolution equations  (\ref{eq:quantumphi})--(\ref{eq:H}) for the 2PI $1/N$ expansion to NLO. 
The original homogeneous field and correlation functions are obtained from averaging the corresponding inhomogeneous ones over some local volume $V$. In general, this "coarse graining" volume $V$ is smaller than the system size $L^d$ and is determined by the statistical error one wants to achieve. It can also be convenient to implement the averaging in Fourier space, which we employ below.

In the following we show that the inhomogeneous 2PI approach contains the classical-statistical simulation results of Sec.~\ref{sec:classical_statistical_simulation} already at leading order (LO) in the large-$N$ expansion. The equations are given by (\ref{eq:quantumphi})--(\ref{eq:mass_squared}) with $\Sigma^F = \Sigma^\rho = 0$ at LO and read
\begin{eqnarray}  && \!\!\!\! \left( \partial_t^2+\gamma\partial_t-\partial_x^2 + \mu^2+\frac{\lambda}{6N} \left[ F_{cc}(t,t;x,x)+ \phi^2(t,x) \right] \right)\phi_a(t,x) = J_a(t,x) \, , 
\nonumber\\ && \!\!\!\! \left(\partial_t^2+\gamma\partial_t-\partial_x^2+\mu^2+\frac{\lambda}{6N}\left[F_{cc}(t,t;x,x) + \phi^2(t,x) \right] \right)F_{ab}(t,t';x,y) = 0 \, ,
\nonumber \\ && \!\!\!\! \left(\partial_t^2+\gamma\partial_t-\partial_x^2+\mu^2+\frac{\lambda}{6N}\left[F_{cc}(t,t;x,x) + \phi^2(t,x) \right]\right)\rho_{ab}(t,t';x,y) = 0 \, .
\nonumber \\ \label{eq:phiFrhoLO}
\end{eqnarray}     
One observes that the dynamics for the spectral function $\rho$ does not influence the evolution of the field $\phi$ and the statistical function $F$ at this order. We will not consider it in the following. Related dynamics based on an inhomogeneous Hartree approximation have been considered in Ref.~\cite{Salle:2002fu}.

The inhomogeneous evolution equations (\ref{eq:phiFrhoLO}) have to be supplemented by initial conditions, which we formulate here in spatial Fourier space using the functions
\begin{equation}
\chi_a(t=0,p) \, = \, \frac{1}{\sqrt{2 \omega_p}}\, e^{i \alpha_a(p)} \quad , \quad \partial_t\chi_a(t,p)|_{t=0} \, = \, \sqrt{\frac{\omega_p}{2}}\, e^{i \beta_a(p)} \, ,
\end{equation}
again all in units of appropriate powers of $a_s$, with random numbers $\alpha_a(p)$ and $\beta_a(p)$. From this we compute the Fourier transform to obtain $\chi_a(t=0,x)$ and derivatives. The initial correlation functions are given by (\ref{eq:initialrho}) for the spectral function and for the statistical function we take
\begin{eqnarray}
F_{ab}(t,t';x,y)_{t=t'=0} & = & \chi_a(0,x) \chi_b(0,y) \, , \nonumber\\
\partial_t F_{ab}(t,t';x,y)_{t=t'=0} & = & 0 \, , \label{eq:initialinh}\\
\partial_t \partial_{t'}F_{ab}(t,t';x,y)_{t=t'=0} & = & \partial_t \chi_a(t,x)_{t=0} \partial_{t'}\chi_b(t',y)_{t'=0} \, , \nonumber
\end{eqnarray}
from which we also determine the initial spatial derivatives. The initial field and also its initial derivatives are taken to vanish with  
\begin{equation}
\phi_{a}(t=0,x) \, \equiv \, 0 \, .
\label{eq:fieldinitialinh}
\end{equation}
This corresponds to the Gaussian initial conditions for averaged quantities as employed in (\ref{eq:fieldinitialcl}) and (\ref{eq:field_inital_conditions}), or (\ref{eq:initialrho})--(\ref{eq:fieldinitial}) of previous sections. 

We consider $J = 0$ for which the field vanishes at all times, $\phi_a(t,x) = 0$, with the zero initial field (\ref{eq:fieldinitialinh}) and derivatives. Using (\ref{eq:initialinh}) one directly verifies the equivalence of the LO evolution equation for $F$ with the classical field equation (\ref{eq:equation_of_motion}) if the latter is multiplied by one power of the field. As a consequence, at LO the dynamics is purely classical. It should be emphasized that the initial conditions are the same only on average. Therefore, the differences concern the inhomogeneous initial conditions and the averaging procedure to obtain the homogeneous correlation functions: While in the classical-statistical simulations of Sec.~\ref{sec:classical_statistical_simulation} the average is done over many individual runs with inhomogeneous fields, here the result is obtained from a single integration of the differential equation for $F_{ab}(t,t';x,y)$ and a volume average. In this case, the comparison between the classical-statistical dynamics of Sec.~\ref{sec:classical_statistical_simulation} and the present LO dynamics boils down to a comparison between ensemble and (local) volume averages.  

\begin{figure}[t]
\begin{center}
\input{inhom.tex}
\caption[inhomogeneous 2PI to LO]{Comparison of results for the statistical two-point function $F$ as a function of the momentum at $t=40$. The results from the inhomogeneous 2PI large-$N$ expansion at LO (symbols) and from classical-statistical simulations (dotted line) agree very well. Shown is also the kink fit (solid line) of Sec.~\ref{sec:classical_statistical_simulation}}
\label{fig:inhom}
\end{center}
\end{figure}
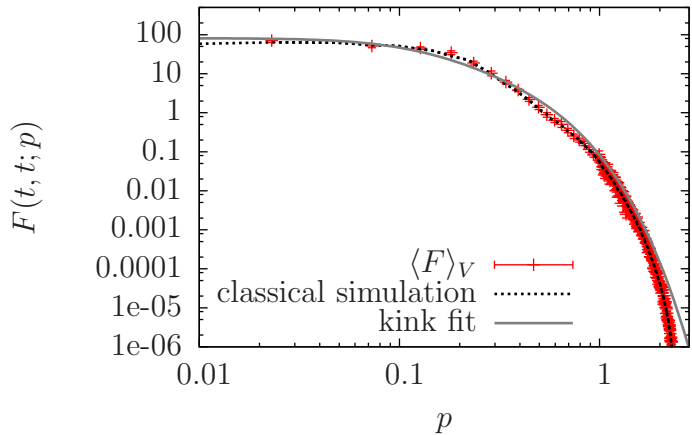
To demonstrate that the volume average of $F(t,t';x,y)$ indeed becomes homogeneous and that the results agree well to those obtained from the classical-statistical simulations of Sec.~\ref{sec:classical_statistical_simulation}, we integrate the evolution equation for $F(t,t';x,y)$ for the $N=1$ component theory in configuration space on a spatial lattice with $L=8192$ and periodic boundary conditions. We then spatially Fourier transform the results to get correlation functions in momentum space, in which we also do the averaging to obtain the homogeneous correlation function. More precisely, we evaluate the inhomogeneous two-point function $F(t,t';p,p')$ at $p'=-p$ and compute the homogeneous average from binning $F(t,t';p,-p)$ for all $p$ belonging to a given bin on a logarithmic scale. In Fig.~\ref{fig:inhom} we show the result using 1024 bins, along with the statistical error. We plot the correlation function as a function of momentum at $t=40$. At this time important contributions from topological defects can be observed according to the discussion of Sec.~\ref{sec:classical_statistical_simulation}. For comparison we also give the previous classical-statistical simulation result and the fit for the analytical defect ansatz. 
The agreement is very good and the statistical errors barely visible from the plot. This demonstrates that the inhomogeneous 2PI approach contains the classical-statistical simulation results of Sec.~\ref{sec:classical_statistical_simulation} already at LO in the large-$N$ expansion.

At NLO in the 2PI $1/N$ expansion genuine quantum corrections enter the dynamics. Here it is important to note that in this approach, classical-statistical fluctuations and dynamics is included exactly already at LO, whereas quantum corrections are treated in an approximate, truncated form. Though the direct integration of the corresponding NLO equations (\ref{eq:quantumphi})--(\ref{eq:H}) along the same lines is straightforward in principle, it is rather costly numerically, in particular in higher dimensions such as $d=3$. We defer the discussion of suitable simplifications, which allow one to describe defect formation in phenomenologically relevant quantum field theories in practice, to a subsequent publication~\cite{BR}.

\section{Conclusion}
\label{sec:conclusion}

Much of our theoretical knowledge about the dynamics in complex physical situations such as defect formation, nonequilibrium instabilities or turbulence relies on classical simulations. Classical-statistical simulations are based on inhomogeneous field dynamics and importance sampling to describe homogeneous systems. In many cases these are known to be reliable approximations for quantum systems, at least for short times before the approach to thermal equilibrium characterized by Bose-Einstein or Fermi-Dirac distributions. 

In this work we have proposed a quantum approach which combines the successful aspects of inhomogeneous classical simulations with the ability to take into account quantum corrections. This is necessary since standard nonperturbative quantum approaches, such as the 2PI large-$N$ expansion for homogeneous fields, turn out to be not appropriate for a quantitative description of defects for one- and two-component theories at low expansion orders.
We have demonstrated that the result of Ref.~\cite{Rajantie:2006gy} is correct, and that the discrepancy is indeed qualitative rather than quantitative. To see this we considered the procedure of introducing a source to the equation and continuously bringing the source to zero. At some threshold value of $J$, the potential acquires degenerate minima, thus allowing for defects, and this is the point where the discrepancy is seen to kick in. Introducing a non-zero $J$ also leads to including a non-zero field expectation value, which is necessary to follow the true field value in a symmetry breaking phase transition. The omission of this in Ref.~\cite{Rajantie:2006gy} was one of the main motivations to start this investigation. In general, we find that NLO results for $N=2$ agree somewhat better to those obtained from classical-statistical simulations than for $N=1$ as expected for a large-$N$ expansion. It would be very interesting to extend the study to NNLO, however, a more appropriate calculation should be based on the 2PI effective action for inhomogeneous fields and a volume average. We emphasize that the latter does not involve any double counting which could appear in sampling prescriptions for inhomogeneous quantum evolutions. We have shown that already at leading order in a $1/N$ expansion the results from classical-statistical simulations are recovered. At NLO quantum corrections enter the dynamics. This provides a clear procedure to describe defect formation in quantum field theory, though it is numerically very costly. The computations of the inhomogeneous NLO approximation along with a discussion of suitable simplifications will be presented in a subsequent publication \cite{BR}. 

A further promising application of our method concerns the question of thermalization in quantum chromodynamics in the context of heavy-ion collisions. Classical-statistical simulations in lattice gauge theory have been applied to plasma instabilities and the question of isotropization at early times \cite{Romatschke:2005pm,Berges:2007re}. At low momenta the instability dynamics is expected to be well approximated by classical methods. However, at higher momenta quantum corrections may play an important role, because occupation numbers are low at large momenta. A 2PI loop expansion for inhomogeneous gauge fields along the lines of this work would include the classical-statistical simulation results already at lowest order. Going beyond lowest order in a 2PI loop expansion may be an appropriate way to include relevant quantum corrections at higher momenta.\\

This work is supported in part by the BMBF grant 06DA9018.

\appendix

\section{Calculation of the 2PI effective action}
\label{app:Calculation of the 2PI action}

The classical field equation (\ref{eq:equation_of_motion}) contains a damping term, which can be obtained from a principle of least action if it is viewed as the equation of motion for an undamped 2+1 dimensional scalar field theory in an expanding anisotropic geometry with metric \cite{Rajantie:2006gy}
\begin{equation}
    ds^2=-dt^2+dx^2+a^2(t)dy^2 \quad ,\quad a(t)=a_0\exp(\gamma t) \, .
\end{equation}
Then \eqref{eq:equation_of_motion} follows from the equation of motion for the field $\phi(t,x,y)$,
\begin{equation} \partial_t^2\phi_a+\frac{\dot{a}}{a}\partial_t\phi_a-\partial_x^2\phi_a-\frac{1}{a^2}\partial_y^2\phi_a+\mu^2\phi_a+\frac{\lambda}{6N}(\phi_b\phi_b)\phi_a-J_a=0
\label{eq:fieldexpanding}
\end{equation}
by neglecting the dependence of the field on the spatial variable y. 

\subsection{2PI effective action in expanding geometry}

To obtain a quantum description of our model, we will provide some details of the calculation of the 2PI effective action in an expanding geometry in this section. For similar discussions see Refs.~\cite{Tranberg:2008ae,Calzetta:1986cq}. To keep the formulation general, we consider an arbitrary metric $g^{\mu\nu}$ with signature $(-++\dots)$. For a compact notation variables as $x=(x^0,\bx)$ are taken to denote space-time variables in $d+1$ dimension, in contrast to the convention in the rest of this paper. At some points we insert the special metric of our model to recover the equations of Sec.~\ref{sec:2PI_effective_action}. For a general metric $g^{\mu\nu}$ the classical action reads
\begin{equation}
S[\varphi;g] = \int_x\sqrt{-g(x)}\left(-\frac{1}{2}g^{\mu\nu}(x)\partial_{\mu}\varphi_a(x)\partial_{\nu}\varphi_a(x)-V(\varphi^2)\right) \, .
\end{equation}
Here $\int_x\equiv\int_{\mathscr{C}}dx^0\int d^dx$ involves integration over the Schwinger-Keldysh closed time contour $\mathscr{C}$ \cite{Schwinger:1960qe}. The determinant of the metric tensor is denoted as $g=\det g^{\mu\nu}$. For our model discussed in this work we have 
\begin{equation}
\sqrt{-g(x)}=a(x^0)=a_0\exp{(\gamma x^0)} \, . 
\end{equation}
The inverse classical propagator is given by
\begin{eqnarray}\label{eq:free_propagator}
    iG_{0,ab}^{-1}(x,y;\varphi) & = & \frac{1}{\sqrt{-g(x)}}\frac{1}{\sqrt{-g(y)}}\frac{\delta^2S[\varphi;g]}{\delta\varphi_a(x)\delta\varphi_b(y)} \nonumber\\
    & = & - D_{ab}(x;\varphi) \frac{\delta_{\mathscr{C}}(x-y)}{\sqrt{-g(y)}} \, ,
\end{eqnarray}
where for our model $D$ reads
\begin{eqnarray}
    D_{ab}(x;\varphi) & = &  \left(\partial_{x^0}^2+\frac{\dot{a}}{a}\partial_{x^0}-\partial_{\bx}^2+\mu^2+\frac{\lambda}{6N}(\varphi_c(x)\varphi_c(x))\right)\delta_{ab} \nonumber\\
  &&  +\frac{\lambda}{3N}\varphi_a(x)\varphi_b(x) .
\end{eqnarray}
Introducing sources $J_a(x)$ and $R_{ab}(x,y)$ we construct the generating functional  
\begin{eqnarray}
    Z[J,R;g] & = & \int\mathscr{D}\varphi \exp\bigg\{i\bigg(S[\varphi;g]+\int_x\sqrt{-g(x)}J_a(x)\varphi_a(x)\notag\\
    & & +\frac{1}{2}\int_{x,y}\sqrt{-g(x)}\sqrt{-g(y)}R_{ab}(x,y)\varphi_a(x)\varphi_b(y)\bigg)\bigg\} \, , \\
    W[J,R;g] & = & -i\ln Z[J,R;g] \, ,
\end{eqnarray}
where $W[J,R;g]$ is the generating functional for connected Green's functions.
The field expectation value $\phi_a$ and the connected two-point function $G_{ab}$ are defined as 
\begin{eqnarray}\label{eq:variation_w}
    \frac{1}{\sqrt{-g(x)}}\frac{\delta W[J,R;g]}{\delta J_a(x)} & = & \phi_a(x)\notag \, , \\
    \frac{1}{\sqrt{-g(x)}}\frac{1}{\sqrt{-g(y)}}\frac{\delta W[J,R;g]}{\delta R_{ab}(x,y)} & = & \frac{1}{2}\left[\phi_a(x)\phi_b(y)+G_{ab}(x,y)\right] \, .
\end{eqnarray}
The 2PI effective action $\Gamma$ is the Legendre transform, 
\begin{eqnarray}\label{eq:2PI_action}
    \Gamma[\phi,G;g] & = & W[J,R;g]-\int_x\frac{\delta W[J,R;g]}{\delta J_a(x)}J_a(x)-\int_{x,y}\frac{\delta W[J,R;g]}{\delta R_{ab}(x,y)}R_{ab}(x,y)\notag\\
    & = & W[J,R;g]-\int_x\sqrt{-g(x)}\phi_a(x)J_a(x) -\frac{1}{2}\int_{x,y}\sqrt{-g(x)}\sqrt{-g(y)} \notag\\
    & & \times R_{ab}(x,y)\left[\phi_a(x)\phi_b(y)
    + G_{ab}(x,y) \right] \, ,
\end{eqnarray}
which can be written as~\cite{Cornwall:1974vz}
\begin{equation}
  \Gamma[\phi,G;g] = S[\phi;g]+\frac{i}{2}{\rm{Tr}}\ln G^{-1}+\frac{i}{2}{\rm{Tr}}G_0^{-1}G+\Gamma_2[\phi,G;g]+\text{const.} \, ,
\end{equation}
where $\Gamma_2$ contains all two-particle irreducible contributions. The equations of motion for the field expectation value and the propagator are
\begin{eqnarray}\label{eq:field_equation_of_motion}
    \frac{1}{\sqrt{-g(x)}}\frac{\delta\Gamma[\phi,G;g]}{\delta\phi_a(x)} & = & -J_a(x)-\int_y\sqrt{-g(y)}R_{ab}(x,y)\phi_b(y) \, , \nonumber\\
    \frac{1}{\sqrt{-g(x)}}\frac{1}{\sqrt{-g(y)}}\frac{\delta\Gamma[\phi,G;g]}{\delta G_{ab}(x,y)} & = & -\frac{1}{2}R_{ab}(x,y) \, .
\end{eqnarray}
For $R_{ab}(x,y) = 0$ the second equation corresponds to the Schwinger-Dyson equation
\begin{equation}\label{eq:DSE}
 G_{ab}^{-1}(x,y)=G_{0\;ab}^{-1}(x,y;\phi)-\Sigma_{ab}(x,y) \, ,
\end{equation}
where the self-energy $\Sigma$ is defined by
\begin{eqnarray}
 \Sigma_{ab}(x,y) & = & 2i\frac{1}{\sqrt{-g(x)}}\frac{1}{\sqrt{-g(y)}}\frac{\delta\Gamma_2[\phi,G;g]}{\delta G_{ab}(c,y)}\notag\\
 & = & -i\Sigma^{(0)}_{ab}(x)\frac{\delta_{\mathscr{C}}(x-y)}{\sqrt{-g(x)}}+\overline{\Sigma}_{ab}(x,y) 
 \label{eq:sigma}
\end{eqnarray}
In the last equation we have separated a local part $\Sigma^{(0)}_{ab}(x)$ from the self-energy, which for our model is given by
\begin{equation}
\Sigma^{(0)}_{ab}(x) = \frac{\lambda}{6N}F_{cc}(x,x)\delta_{ab} + \frac{\lambda}{3N}F_{ab}(x,x)\, .
\end{equation}
Multiplication of \eqref{eq:DSE} with $G$ and integration leads to the differential evolution equation for the propagator
\begin{eqnarray}\label{eq:evolution equation}
 \lefteqn{\left[D_{ab}^2(x;\phi)+\Sigma^{(0)}_{ab}(x)\right]G_{bc}(x,y)} \nonumber\\
  & = & -i\int_z\sqrt{-g(z)}\overline{\Sigma}_{ab}(x,z)G_{bc}(z,y)-i\delta_{ac}\frac{\delta_{\mathscr{C}}(x-y)}{\sqrt{-g(y)}} \, .
\end{eqnarray}

\subsection[2PI 1/N expansion to NLO]{2PI $1/N$ expansion to NLO}

For the $O(N)$-symmetric scalar field theory with classical potential (\ref{eq:classicalV}) we classify the contributions to the 2PI action by their scaling in powers of $1/N$: 
\begin{equation} \Gamma_2[\phi,G;g]=\Gamma_2^{\text{LO}}[G;g]+\Gamma_2^{\text{NLO}}[\phi,G;g]+\dots.
\end{equation}
The leading order contributions (LO) scale as $N^1$, the next-to-leading order terms (NLO) by $N^0$ and so on. According to Ref.~\cite{Berges:2001fi} at LO this is given by
\begin{equation}
 \Gamma_2^{\text{LO}}[G;g]=-\frac{\lambda}{4!N}\int_x\sqrt{-g(x)}G_{aa}(x,x)G_{bb}(x,x) \,
\end{equation}
and the NLO contributions are
\begin{eqnarray}
 \Gamma_2^{\text{NLO}}[\phi,G;g] & = & \frac{i}{2}{\rm{Tr}}\ln[B(G)] \nonumber \\
 & + & \frac{i\lambda}{6N}\int_{x,y}\sqrt{-g(x)}\sqrt{-g(y)}I(x,y)\phi_a(x)G_{ab}(x,y)\phi_b(y)
\quad \end{eqnarray} 
with the definition
\begin{equation} 
 B(x,y)  =  \frac{\delta_{\mathscr{C}}(x-y)}{\sqrt{-g(x)}\sqrt{-g(y)}}+i\frac{\lambda}{6N}G_{ab}(x,y)G_{ab}(x,y)
\end{equation} 
and the integral equation
\begin{eqnarray}  
 I(x,y) & = & \frac{\lambda}{6N}G_{ab}(x,y)G_{ab}(x,y)\notag\\
 & - & i\frac{\lambda}{6N}\int_z\sqrt{-g(z)}I(x,z)G_{ab}(z,y)G_{ab}(z,y)\, .
\end{eqnarray}

In order to write the evolution equation (\ref{eq:evolution equation}) in a form which is suitable for numerical computations, we identically decompose the time-ordered propagator in a statistical part $F$ and a spectral part $\rho$~\cite{Berges:2004yj},
\begin{equation}\label{eq:decomposition_of_g}
    G_{ab}(x,y) = F_{ab}(x,y)-\frac{i}{2}{\rm{sgn}}_{\mathscr{C}}(x^0-y^0)\rho_{ab}(x,y) \, .
\end{equation}
Similarly, for the non-local part of the self-energy defined in (\ref{eq:sigma}) we employ
\begin{equation}\label{eq:decomposition_of_sigma} \overline{\Sigma}_{ab}(x,y)=\Sigma^F_{ab}(x,y)-\frac{i}{2}{\rm{sgn}}_{\mathscr{C}}(x^0-y^0)\Sigma^{\rho}_{ab}(x,y).
\end{equation}
Applying this decomposition to the evolution equation \eqref{eq:evolution equation} and separating the real and the imaginary part leads to the evolution equations for $F$ and $\rho$
\begin{eqnarray}
    \lefteqn{\left[D_{ab}^2(x;\phi)+\Sigma^{(0)}_{ab}(x)\right]F_{bc}(x,y)}\notag\\
    & = & -\int_{0}^{x^0}dz^0\int d^dz\sqrt{-g(z)}\Sigma^{\rho}_{ab}(x,z)F_{bc}(z,y)\notag\\
    & & +\int_{0}^{y^0}dz^0\int d^dz\sqrt{-g(z)}\Sigma^F_{ab}(x,z)\rho_{bc}(z,y) \, ,\notag\\
    \lefteqn{\left[D_{ab}^2(x;\phi)+\Sigma^{(0)}_{ab}(x)\right]\rho_{bc}(x,y)}\notag\\
    & = & -\int_{y^0}^{x^0}dz^0\int d^dz\sqrt{-g(z)}\Sigma^{\rho}_{ab}(x,z)\rho_{bc}(z,y) \, .
\end{eqnarray}
From equation \eqref{eq:field_equation_of_motion} with vanishing quadratic source $R$ we can derive the corresponding evolution equation for the field,
\begin{eqnarray} \lefteqn{\left[D_{ab}^2(x;\phi=0)+\frac{\lambda}{6N}\phi^2(x)\delta_{ab}+\Sigma^{(0)}_{ab}(x)\right]\phi_b(x)-J_a(x)}\notag\\
    & = & -\int_{0}^{x^0}dy^0\int d^dy\sqrt{-g(y)}\Sigma^{\rho}_{ab}(x,y;\phi=0)\phi_b(y) \, .
\end{eqnarray}

\subsection{Energy-momentum tensor}

The energy-momentum tensor can be calculated by varying the 2PI effective action with respect to $g^{\mu\nu}(x)$,
\begin{equation}
    T_{\mu\nu}(x)=-\frac{2}{\sqrt{-g(x)}}\frac{\delta\Gamma[\phi,G;g]}{\delta g^{\mu\nu}(x)}.
\end{equation}
At NLO in the 2PI $1/N$ expansion this gives  
\begin{eqnarray}
    \lefteqn{T_{\mu\nu}(x)
    \, = \, \partial_{\mu}\phi_a(x)\partial_{\nu}\phi_a(x)}\notag\\
    & & -g_{\mu\nu}(x)\left[\frac{1}{2}g^{\kappa\lambda}\partial_{\kappa}\phi_a(x)\partial_{\lambda}\phi_a(x)+\frac{1}{2}\mu^2\phi_a(x)\phi_a(x)+\frac{\lambda}{4!N}\left(\phi_a(x)\phi_a(x)\right)^2\right]\notag\\
    & & +\left[\partial_{\mu}^x\partial_{\nu}^{x'}F_{aa}(x,x')-\frac{1}{2}g_{\mu\nu}\left(g^{\kappa\lambda}\partial_{\kappa}^x\partial_{\lambda}^{x'}\delta_{ab}+M^2_{ab}(\phi;x)\right)F_{ab}(x,x')\right]\bigg|_{x'=x}\notag\\
    & & -g_{\mu\nu}(x)\frac{\lambda}{4!N}\left[F_{aa}(x,x)\right]^2-g_{\mu\nu}(x)\frac{1}{2}I_F(x,x)\notag\\
    & & -g_{\mu\nu}(x)\frac{1}{2}\left[P_F(x,x)+\frac{\lambda}{3N}H_F(x,x)\right] \, ,
\end{eqnarray}
with the summation functions $I_F$, $P_F$ and $H_F$ as defined in Sec.~\ref{sec:2PI_effective_action}.

\end{document}

%% file: average_phi.tex
\begin{picture}(0,0)%
\includegraphics{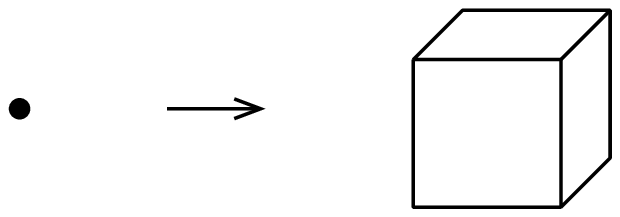}%
\end{picture}%
\setlength{\unitlength}{4144sp}%
\begingroup\makeatletter\ifx\SetFigFont\undefined%
\gdef\SetFigFont#1#2#3#4#5{%
  \reset@font\fontsize{#1}{#2pt}%
  \fontfamily{#3}\fontseries{#4}\fontshape{#5}%
  \selectfont}%
\fi\endgroup%
\begin{picture}(2962,2106)(211,-1390)
\put(226,389){\makebox(0,0)[lb]{\smash{{\SetFigFont{20}{24.0}{\familydefault}{\mddefault}{\updefault}{\color[rgb]{0,0,0}$\phi(t)$}%
}}}}
\put(2476,-1276){\makebox(0,0)[lb]{\smash{{\SetFigFont{20}{24.0}{\familydefault}{\mddefault}{\updefault}{\color[rgb]{0,0,0}$V$}%
}}}}
\put(2026,389){\makebox(0,0)[lb]{\smash{{\SetFigFont{20}{24.0}{\familydefault}{\mddefault}{\updefault}{\color[rgb]{0,0,0}$\phi(t,\vec{x})\Big|_{\langle\phi(t,\vec{x})\rangle_V=\phi(t)}$}%
}}}}
\end{picture}%

%% file: field_analytic_potential.tex
\begingroup
  \makeatletter
  \providecommand\color[2][]{%
    \GenericError{(gnuplot) \space\space\space\@spaces}{%
      Package color not loaded in conjunction with
      terminal option `colourtext'%
    }{See the gnuplot documentation for explanation.%
    }{Either use 'blacktext' in gnuplot or load the package
      color.sty in LaTeX.}%
    \renewcommand\color[2][]{}%
  }%
  \providecommand\includegraphics[2][]{%
    \GenericError{(gnuplot) \space\space\space\@spaces}{%
      Package graphicx or graphics not loaded%
    }{See the gnuplot documentation for explanation.%
    }{The gnuplot epslatex terminal needs graphicx.sty or graphics.sty.}%
    \renewcommand\includegraphics[2][]{}%
  }%
  \providecommand\rotatebox[2]{#2}%
  \@ifundefined{ifGPcolor}{%
    \newif\ifGPcolor
    \GPcolortrue
  }{}%
  \@ifundefined{ifGPblacktext}{%
    \newif\ifGPblacktext
    \GPblacktexttrue
  }{}%
  \let\gplgaddtomacro\g@addto@macro
  \gdef\gplbacktext{}%
  \gdef\gplfronttext{}%
  \makeatother
  \ifGPblacktext
    \def\colorrgb#1{}%
    \def\colorgray#1{}%
  \else
    \ifGPcolor
      \def\colorrgb#1{\color[rgb]{#1}}%
      \def\colorgray#1{\color[gray]{#1}}%
      \expandafter\def\csname LTw\endcsname{\color{white}}%
      \expandafter\def\csname LTb\endcsname{\color{black}}%
      \expandafter\def\csname LTa\endcsname{\color{black}}%
      \expandafter\def\csname LT0\endcsname{\color[rgb]{1,0,0}}%
      \expandafter\def\csname LT1\endcsname{\color[rgb]{0,1,0}}%
      \expandafter\def\csname LT2\endcsname{\color[rgb]{0,0,1}}%
      \expandafter\def\csname LT3\endcsname{\color[rgb]{1,0,1}}%
      \expandafter\def\csname LT4\endcsname{\color[rgb]{0,1,1}}%
      \expandafter\def\csname LT5\endcsname{\color[rgb]{1,1,0}}%
      \expandafter\def\csname LT6\endcsname{\color[rgb]{0,0,0}}%
      \expandafter\def\csname LT7\endcsname{\color[rgb]{1,0.3,0}}%
      \expandafter\def\csname LT8\endcsname{\color[rgb]{0.5,0.5,0.5}}%
    \else
      \def\colorrgb#1{\color{black}}%
      \def\colorgray#1{\color[gray]{#1}}%
      \expandafter\def\csname LTw\endcsname{\color{white}}%
      \expandafter\def\csname LTb\endcsname{\color{black}}%
      \expandafter\def\csname LTa\endcsname{\color{black}}%
      \expandafter\def\csname LT0\endcsname{\color{black}}%
      \expandafter\def\csname LT1\endcsname{\color{black}}%
      \expandafter\def\csname LT2\endcsname{\color{black}}%
      \expandafter\def\csname LT3\endcsname{\color{black}}%
      \expandafter\def\csname LT4\endcsname{\color{black}}%
      \expandafter\def\csname LT5\endcsname{\color{black}}%
      \expandafter\def\csname LT6\endcsname{\color{black}}%
      \expandafter\def\csname LT7\endcsname{\color{black}}%
      \expandafter\def\csname LT8\endcsname{\color{black}}%
    \fi
  \fi
  \setlength{\unitlength}{0.0500bp}%
  \begin{picture}(5760.00,3528.00)%
    \gplgaddtomacro\gplbacktext{%
      \csname LTb\endcsname%
      \put(946,704){\makebox(0,0)[r]{\strut{}-4}}%
      \put(946,988){\makebox(0,0)[r]{\strut{}-3}}%
      \put(946,1273){\makebox(0,0)[r]{\strut{}-2}}%
      \put(946,1557){\makebox(0,0)[r]{\strut{}-1}}%
      \put(946,1842){\makebox(0,0)[r]{\strut{} 0}}%
      \put(946,2126){\makebox(0,0)[r]{\strut{} 1}}%
      \put(946,2411){\makebox(0,0)[r]{\strut{} 2}}%
      \put(946,2695){\makebox(0,0)[r]{\strut{} 3}}%
      \put(946,2980){\makebox(0,0)[r]{\strut{} 4}}%
      \put(946,3264){\makebox(0,0)[r]{\strut{} 5}}%
      \put(1078,484){\makebox(0,0){\strut{}-4}}%
      \put(1622,484){\makebox(0,0){\strut{}-3}}%
      \put(2166,484){\makebox(0,0){\strut{}-2}}%
      \put(2710,484){\makebox(0,0){\strut{}-1}}%
      \put(3254,484){\makebox(0,0){\strut{} 0}}%
      \put(3798,484){\makebox(0,0){\strut{} 1}}%
      \put(4342,484){\makebox(0,0){\strut{} 2}}%
      \put(4886,484){\makebox(0,0){\strut{} 3}}%
      \put(5430,484){\makebox(0,0){\strut{} 4}}%
      \put(440,1984){\rotatebox{90}{\makebox(0,0){\strut{}$V(\varphi)$}}}%
      \put(3254,154){\makebox(0,0){\strut{}$\varphi$}}%
    }%
    \gplgaddtomacro\gplfronttext{%
      \csname LTb\endcsname%
      \put(4443,3091){\makebox(0,0)[r]{\strut{}J=0}}%
      \csname LTb\endcsname%
      \put(4443,2871){\makebox(0,0)[r]{\strut{}J=0.4}}%
      \csname LTb\endcsname%
      \put(4443,2651){\makebox(0,0)[r]{\strut{}J=1}}%
    }%
    \gplbacktext
    \put(0,0){\includegraphics{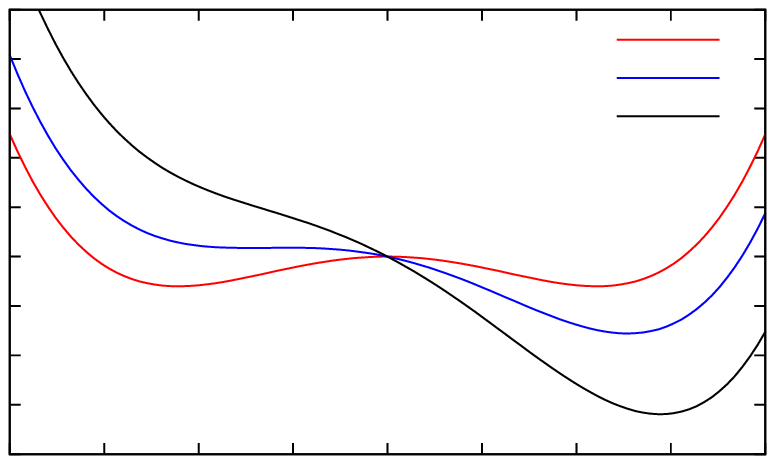}}%
    \gplfronttext
  \end{picture}%
\endgroup

%% file: symmetry_breaking.tex
\begingroup
  \makeatletter
  \providecommand\color[2][]{%
    \GenericError{(gnuplot) \space\space\space\@spaces}{%
      Package color not loaded in conjunction with
      terminal option `colourtext'%
    }{See the gnuplot documentation for explanation.%
    }{Either use 'blacktext' in gnuplot or load the package
      color.sty in LaTeX.}%
    \renewcommand\color[2][]{}%
  }%
  \providecommand\includegraphics[2][]{%
    \GenericError{(gnuplot) \space\space\space\@spaces}{%
      Package graphicx or graphics not loaded%
    }{See the gnuplot documentation for explanation.%
    }{The gnuplot epslatex terminal needs graphicx.sty or graphics.sty.}%
    \renewcommand\includegraphics[2][]{}%
  }%
  \providecommand\rotatebox[2]{#2}%
  \@ifundefined{ifGPcolor}{%
    \newif\ifGPcolor
    \GPcolortrue
  }{}%
  \@ifundefined{ifGPblacktext}{%
    \newif\ifGPblacktext
    \GPblacktexttrue
  }{}%
  \let\gplgaddtomacro\g@addto@macro
  \gdef\gplbacktext{}%
  \gdef\gplfronttext{}%
  \makeatother
  \ifGPblacktext
    \def\colorrgb#1{}%
    \def\colorgray#1{}%
  \else
    \ifGPcolor
      \def\colorrgb#1{\color[rgb]{#1}}%
      \def\colorgray#1{\color[gray]{#1}}%
      \expandafter\def\csname LTw\endcsname{\color{white}}%
      \expandafter\def\csname LTb\endcsname{\color{black}}%
      \expandafter\def\csname LTa\endcsname{\color{black}}%
      \expandafter\def\csname LT0\endcsname{\color[rgb]{1,0,0}}%
      \expandafter\def\csname LT1\endcsname{\color[rgb]{0,1,0}}%
      \expandafter\def\csname LT2\endcsname{\color[rgb]{0,0,1}}%
      \expandafter\def\csname LT3\endcsname{\color[rgb]{1,0,1}}%
      \expandafter\def\csname LT4\endcsname{\color[rgb]{0,1,1}}%
      \expandafter\def\csname LT5\endcsname{\color[rgb]{1,1,0}}%
      \expandafter\def\csname LT6\endcsname{\color[rgb]{0,0,0}}%
      \expandafter\def\csname LT7\endcsname{\color[rgb]{1,0.3,0}}%
      \expandafter\def\csname LT8\endcsname{\color[rgb]{0.5,0.5,0.5}}%
    \else
      \def\colorrgb#1{\color{black}}%
      \def\colorgray#1{\color[gray]{#1}}%
      \expandafter\def\csname LTw\endcsname{\color{white}}%
      \expandafter\def\csname LTb\endcsname{\color{black}}%
      \expandafter\def\csname LTa\endcsname{\color{black}}%
      \expandafter\def\csname LT0\endcsname{\color{black}}%
      \expandafter\def\csname LT1\endcsname{\color{black}}%
      \expandafter\def\csname LT2\endcsname{\color{black}}%
      \expandafter\def\csname LT3\endcsname{\color{black}}%
      \expandafter\def\csname LT4\endcsname{\color{black}}%
      \expandafter\def\csname LT5\endcsname{\color{black}}%
      \expandafter\def\csname LT6\endcsname{\color{black}}%
      \expandafter\def\csname LT7\endcsname{\color{black}}%
      \expandafter\def\csname LT8\endcsname{\color{black}}%
    \fi
  \fi
  \setlength{\unitlength}{0.0500bp}%
  \begin{picture}(5760.00,3528.00)%
    \gplgaddtomacro\gplbacktext{%
      \csname LTb\endcsname%
      \put(1210,704){\makebox(0,0)[r]{\strut{} 0}}%
      \put(1210,1131){\makebox(0,0)[r]{\strut{} 0.5}}%
      \put(1210,1557){\makebox(0,0)[r]{\strut{} 1}}%
      \put(1210,1984){\makebox(0,0)[r]{\strut{} 1.5}}%
      \put(1210,2411){\makebox(0,0)[r]{\strut{} 2}}%
      \put(1210,2837){\makebox(0,0)[r]{\strut{} 2.5}}%
      \put(1210,3264){\makebox(0,0)[r]{\strut{} 3}}%
      \put(1342,484){\makebox(0,0){\strut{} 0.001}}%
      \put(2705,484){\makebox(0,0){\strut{} 0.01}}%
      \put(4067,484){\makebox(0,0){\strut{} 0.1}}%
      \put(5430,484){\makebox(0,0){\strut{} 1}}%
      \put(440,1984){\rotatebox{90}{\makebox(0,0){\strut{}$\phi_{\rm{cl}}$}}}%
      \put(3386,154){\makebox(0,0){\strut{}$J$}}%
    }%
    \gplgaddtomacro\gplfronttext{%
      \csname LTb\endcsname%
      \put(4443,1317){\makebox(0,0)[r]{\strut{}classical}}%
      \csname LTb\endcsname%
      \put(4443,1097){\makebox(0,0)[r]{\strut{}$\gamma=0.4$}}%
      \csname LTb\endcsname%
      \put(4443,877){\makebox(0,0)[r]{\strut{}$\gamma=0$}}%
    }%
    \gplbacktext
    \put(0,0){\includegraphics{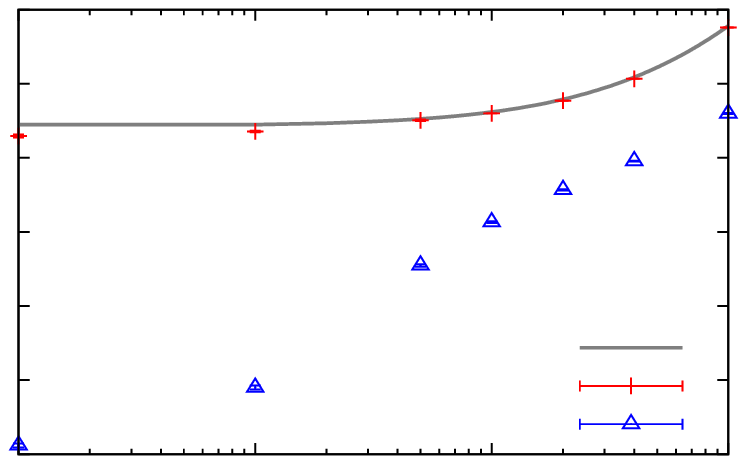}}%
    \gplfronttext
  \end{picture}%
\endgroup

%% file: CLSTAT_phi_F.tex
\begingroup
  \makeatletter
  \providecommand\color[2][]{%
    \GenericError{(gnuplot) \space\space\space\@spaces}{%
      Package color not loaded in conjunction with
      terminal option `colourtext'%
    }{See the gnuplot documentation for explanation.%
    }{Either use 'blacktext' in gnuplot or load the package
      color.sty in LaTeX.}%
    \renewcommand\color[2][]{}%
  }%
  \providecommand\includegraphics[2][]{%
    \GenericError{(gnuplot) \space\space\space\@spaces}{%
      Package graphicx or graphics not loaded%
    }{See the gnuplot documentation for explanation.%
    }{The gnuplot epslatex terminal needs graphicx.sty or graphics.sty.}%
    \renewcommand\includegraphics[2][]{}%
  }%
  \providecommand\rotatebox[2]{#2}%
  \@ifundefined{ifGPcolor}{%
    \newif\ifGPcolor
    \GPcolortrue
  }{}%
  \@ifundefined{ifGPblacktext}{%
    \newif\ifGPblacktext
    \GPblacktexttrue
  }{}%
  \let\gplgaddtomacro\g@addto@macro
  \gdef\gplbacktext{}%
  \gdef\gplfronttext{}%
  \makeatother
  \ifGPblacktext
    \def\colorrgb#1{}%
    \def\colorgray#1{}%
  \else
    \ifGPcolor
      \def\colorrgb#1{\color[rgb]{#1}}%
      \def\colorgray#1{\color[gray]{#1}}%
      \expandafter\def\csname LTw\endcsname{\color{white}}%
      \expandafter\def\csname LTb\endcsname{\color{black}}%
      \expandafter\def\csname LTa\endcsname{\color{black}}%
      \expandafter\def\csname LT0\endcsname{\color[rgb]{1,0,0}}%
      \expandafter\def\csname LT1\endcsname{\color[rgb]{0,1,0}}%
      \expandafter\def\csname LT2\endcsname{\color[rgb]{0,0,1}}%
      \expandafter\def\csname LT3\endcsname{\color[rgb]{1,0,1}}%
      \expandafter\def\csname LT4\endcsname{\color[rgb]{0,1,1}}%
      \expandafter\def\csname LT5\endcsname{\color[rgb]{1,1,0}}%
      \expandafter\def\csname LT6\endcsname{\color[rgb]{0,0,0}}%
      \expandafter\def\csname LT7\endcsname{\color[rgb]{1,0.3,0}}%
      \expandafter\def\csname LT8\endcsname{\color[rgb]{0.5,0.5,0.5}}%
    \else
      \def\colorrgb#1{\color{black}}%
      \def\colorgray#1{\color[gray]{#1}}%
      \expandafter\def\csname LTw\endcsname{\color{white}}%
      \expandafter\def\csname LTb\endcsname{\color{black}}%
      \expandafter\def\csname LTa\endcsname{\color{black}}%
      \expandafter\def\csname LT0\endcsname{\color{black}}%
      \expandafter\def\csname LT1\endcsname{\color{black}}%
      \expandafter\def\csname LT2\endcsname{\color{black}}%
      \expandafter\def\csname LT3\endcsname{\color{black}}%
      \expandafter\def\csname LT4\endcsname{\color{black}}%
      \expandafter\def\csname LT5\endcsname{\color{black}}%
      \expandafter\def\csname LT6\endcsname{\color{black}}%
      \expandafter\def\csname LT7\endcsname{\color{black}}%
      \expandafter\def\csname LT8\endcsname{\color{black}}%
    \fi
  \fi
  \setlength{\unitlength}{0.0500bp}%
  \begin{picture}(7200.00,5040.00)%
    \gplgaddtomacro\gplbacktext{%
      \csname LTb\endcsname%
      \put(1188,880){\makebox(0,0)[r]{\strut{} 0.01}}%
      \put(1188,1543){\makebox(0,0)[r]{\strut{} 0.1}}%
      \put(1188,2206){\makebox(0,0)[r]{\strut{} 1}}%
      \put(1320,660){\makebox(0,0){\strut{} 1}}%
      \put(3148,660){\makebox(0,0){\strut{} 10}}%
      \put(4976,660){\makebox(0,0){\strut{} 100}}%
      \put(6804,660){\makebox(0,0){\strut{} 1000}}%
      \put(286,1775){\rotatebox{90}{\makebox(0,0){\strut{}$\phi_{\rm{cl}}(t)$}}}%
      \put(4062,330){\makebox(0,0){\strut{}$t$}}%
    }%
    \gplgaddtomacro\gplfronttext{%
      \csname LTb\endcsname%
      \put(5817,1713){\makebox(0,0)[r]{\strut{}$J=0$}}%
      \csname LTb\endcsname%
      \put(5817,1493){\makebox(0,0)[r]{\strut{}$J=0.01$}}%
      \csname LTb\endcsname%
      \put(5817,1273){\makebox(0,0)[r]{\strut{}$J=0.04$}}%
      \csname LTb\endcsname%
      \put(5817,1053){\makebox(0,0)[r]{\strut{}$J=0.4$}}%
    }%
    \gplgaddtomacro\gplbacktext{%
      \csname LTb\endcsname%
      \put(1188,2671){\makebox(0,0)[r]{\strut{} 0.01}}%
      \put(1188,3034){\makebox(0,0)[r]{\strut{} 0.1}}%
      \put(1188,3398){\makebox(0,0)[r]{\strut{} 1}}%
      \put(1188,3761){\makebox(0,0)[r]{\strut{} 10}}%
      \put(1188,4125){\makebox(0,0)[r]{\strut{} 100}}%
      \put(1320,2451){\makebox(0,0){\strut{}}}%
      \put(3148,2451){\makebox(0,0){\strut{}}}%
      \put(4976,2451){\makebox(0,0){\strut{}}}%
      \put(6804,2451){\makebox(0,0){\strut{}}}%
      \put(286,3525){\rotatebox{90}{\makebox(0,0){\strut{}$F_{\rm{cl}}(t,t;p)$}}}%
    }%
    \gplgaddtomacro\gplfronttext{%
    }%
    \gplbacktext
    \put(0,0){\includegraphics{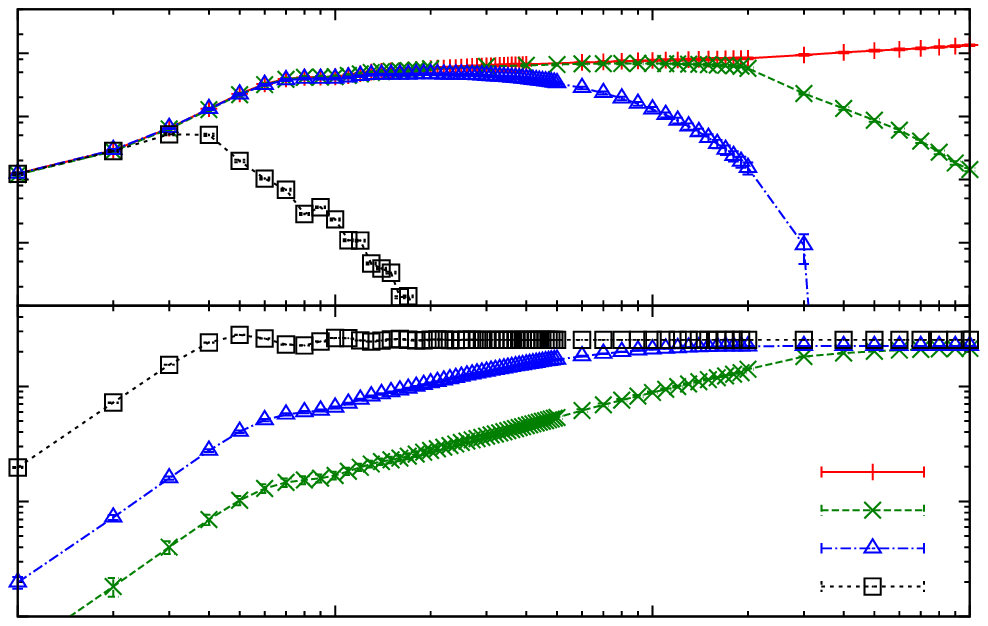}}%
    \gplfronttext
  \end{picture}%
\endgroup

%% file: defects.tex
\begingroup
  \makeatletter
  \providecommand\color[2][]{%
    \GenericError{(gnuplot) \space\space\space\@spaces}{%
      Package color not loaded in conjunction with
      terminal option `colourtext'%
    }{See the gnuplot documentation for explanation.%
    }{Either use 'blacktext' in gnuplot or load the package
      color.sty in LaTeX.}%
    \renewcommand\color[2][]{}%
  }%
  \providecommand\includegraphics[2][]{%
    \GenericError{(gnuplot) \space\space\space\@spaces}{%
      Package graphicx or graphics not loaded%
    }{See the gnuplot documentation for explanation.%
    }{The gnuplot epslatex terminal needs graphicx.sty or graphics.sty.}%
    \renewcommand\includegraphics[2][]{}%
  }%
  \providecommand\rotatebox[2]{#2}%
  \@ifundefined{ifGPcolor}{%
    \newif\ifGPcolor
    \GPcolortrue
  }{}%
  \@ifundefined{ifGPblacktext}{%
    \newif\ifGPblacktext
    \GPblacktexttrue
  }{}%
  \let\gplgaddtomacro\g@addto@macro
  \gdef\gplbacktext{}%
  \gdef\gplfronttext{}%
  \makeatother
  \ifGPblacktext
    \def\colorrgb#1{}%
    \def\colorgray#1{}%
  \else
    \ifGPcolor
      \def\colorrgb#1{\color[rgb]{#1}}%
      \def\colorgray#1{\color[gray]{#1}}%
      \expandafter\def\csname LTw\endcsname{\color{white}}%
      \expandafter\def\csname LTb\endcsname{\color{black}}%
      \expandafter\def\csname LTa\endcsname{\color{black}}%
      \expandafter\def\csname LT0\endcsname{\color[rgb]{1,0,0}}%
      \expandafter\def\csname LT1\endcsname{\color[rgb]{0,1,0}}%
      \expandafter\def\csname LT2\endcsname{\color[rgb]{0,0,1}}%
      \expandafter\def\csname LT3\endcsname{\color[rgb]{1,0,1}}%
      \expandafter\def\csname LT4\endcsname{\color[rgb]{0,1,1}}%
      \expandafter\def\csname LT5\endcsname{\color[rgb]{1,1,0}}%
      \expandafter\def\csname LT6\endcsname{\color[rgb]{0,0,0}}%
      \expandafter\def\csname LT7\endcsname{\color[rgb]{1,0.3,0}}%
      \expandafter\def\csname LT8\endcsname{\color[rgb]{0.5,0.5,0.5}}%
    \else
      \def\colorrgb#1{\color{black}}%
      \def\colorgray#1{\color[gray]{#1}}%
      \expandafter\def\csname LTw\endcsname{\color{white}}%
      \expandafter\def\csname LTb\endcsname{\color{black}}%
      \expandafter\def\csname LTa\endcsname{\color{black}}%
      \expandafter\def\csname LT0\endcsname{\color{black}}%
      \expandafter\def\csname LT1\endcsname{\color{black}}%
      \expandafter\def\csname LT2\endcsname{\color{black}}%
      \expandafter\def\csname LT3\endcsname{\color{black}}%
      \expandafter\def\csname LT4\endcsname{\color{black}}%
      \expandafter\def\csname LT5\endcsname{\color{black}}%
      \expandafter\def\csname LT6\endcsname{\color{black}}%
      \expandafter\def\csname LT7\endcsname{\color{black}}%
      \expandafter\def\csname LT8\endcsname{\color{black}}%
    \fi
  \fi
  \setlength{\unitlength}{0.0500bp}%
  \begin{picture}(7200.00,5040.00)%
    \gplgaddtomacro\gplbacktext{%
      \csname LTb\endcsname%
      \put(1170,3114){\makebox(0,0)[r]{\strut{} 1e-08}}%
      \put(1170,3416){\makebox(0,0)[r]{\strut{} 1e-06}}%
      \put(1170,3718){\makebox(0,0)[r]{\strut{} 0.0001}}%
      \put(1170,4021){\makebox(0,0)[r]{\strut{} 0.01}}%
      \put(1170,4323){\makebox(0,0)[r]{\strut{} 1}}%
      \put(1170,4625){\makebox(0,0)[r]{\strut{} 100}}%
      \put(1302,2894){\makebox(0,0){\strut{} 0.01}}%
      \put(2183,2894){\makebox(0,0){\strut{} 0.1}}%
      \put(3064,2894){\makebox(0,0){\strut{} 1}}%
      \put(268,3945){\rotatebox{90}{\makebox(0,0){\strut{}$F(t,t;p)$}}}%
      \put(2358,2674){\makebox(0,0){\strut{}$p$}}%
      \put(1408,4361){\makebox(0,0)[l]{\strut{}$N=1$}}%
      \put(1408,4144){\makebox(0,0)[l]{\strut{}$\gamma=0.4$}}%
      \put(1408,3928){\makebox(0,0)[l]{\strut{}$t=40$}}%
    }%
    \gplgaddtomacro\gplfronttext{%
      \csname LTb\endcsname%
      \put(2123,3503){\makebox(0,0)[r]{\strut{}class.}}%
      \csname LTb\endcsname%
      \put(2123,3283){\makebox(0,0)[r]{\strut{}kink}}%
    }%
    \gplgaddtomacro\gplbacktext{%
      \csname LTb\endcsname%
      \put(1038,594){\makebox(0,0)[r]{\strut{} 0.001}}%
      \put(1038,871){\makebox(0,0)[r]{\strut{} 0.01}}%
      \put(1038,1148){\makebox(0,0)[r]{\strut{} 0.1}}%
      \put(1038,1425){\makebox(0,0)[r]{\strut{} 1}}%
      \put(1038,1702){\makebox(0,0)[r]{\strut{} 10}}%
      \put(1038,1979){\makebox(0,0)[r]{\strut{} 100}}%
      \put(1038,2256){\makebox(0,0)[r]{\strut{} 1000}}%
      \put(1170,374){\makebox(0,0){\strut{} 0.01}}%
      \put(2106,374){\makebox(0,0){\strut{} 0.1}}%
      \put(3042,374){\makebox(0,0){\strut{} 1}}%
      \put(268,1425){\rotatebox{90}{\makebox(0,0){\strut{}$F(t,t;p)$}}}%
      \put(2292,154){\makebox(0,0){\strut{}$p$}}%
      \put(2449,2090){\makebox(0,0)[l]{\strut{}$N=1$}}%
      \put(2449,1874){\makebox(0,0)[l]{\strut{}$\gamma=0.02$}}%
      \put(2449,1658){\makebox(0,0)[l]{\strut{}$t=1000$}}%
    }%
    \gplgaddtomacro\gplfronttext{%
      \csname LTb\endcsname%
      \put(2096,995){\makebox(0,0)[r]{\strut{}class.}}%
      \csname LTb\endcsname%
      \put(2096,775){\makebox(0,0)[r]{\strut{}kink}}%
    }%
    \gplgaddtomacro\gplbacktext{%
      \csname LTb\endcsname%
      \put(4698,3114){\makebox(0,0)[r]{\strut{} 1e-08}}%
      \put(4698,3416){\makebox(0,0)[r]{\strut{} 1e-06}}%
      \put(4698,3718){\makebox(0,0)[r]{\strut{} 0.0001}}%
      \put(4698,4021){\makebox(0,0)[r]{\strut{} 0.01}}%
      \put(4698,4323){\makebox(0,0)[r]{\strut{} 1}}%
      \put(4698,4625){\makebox(0,0)[r]{\strut{} 100}}%
      \put(4830,2894){\makebox(0,0){\strut{} 0.01}}%
      \put(5711,2894){\makebox(0,0){\strut{} 0.1}}%
      \put(6592,2894){\makebox(0,0){\strut{} 1}}%
      \put(3796,3945){\rotatebox{90}{\makebox(0,0){\strut{}$F(t,t;p)$}}}%
      \put(5886,2674){\makebox(0,0){\strut{}$p$}}%
      \put(4936,4361){\makebox(0,0)[l]{\strut{}$N=2$}}%
      \put(4936,4144){\makebox(0,0)[l]{\strut{}$\gamma=0.4$}}%
      \put(4936,3928){\makebox(0,0)[l]{\strut{}$t=40$}}%
    }%
    \gplgaddtomacro\gplfronttext{%
      \csname LTb\endcsname%
      \put(5651,3503){\makebox(0,0)[r]{\strut{}class.}}%
      \csname LTb\endcsname%
      \put(5651,3283){\makebox(0,0)[r]{\strut{}texture}}%
    }%
    \gplgaddtomacro\gplbacktext{%
      \csname LTb\endcsname%
      \put(4566,594){\makebox(0,0)[r]{\strut{} 0.001}}%
      \put(4566,871){\makebox(0,0)[r]{\strut{} 0.01}}%
      \put(4566,1148){\makebox(0,0)[r]{\strut{} 0.1}}%
      \put(4566,1425){\makebox(0,0)[r]{\strut{} 1}}%
      \put(4566,1702){\makebox(0,0)[r]{\strut{} 10}}%
      \put(4566,1979){\makebox(0,0)[r]{\strut{} 100}}%
      \put(4566,2256){\makebox(0,0)[r]{\strut{} 1000}}%
      \put(4698,374){\makebox(0,0){\strut{} 0.01}}%
      \put(5634,374){\makebox(0,0){\strut{} 0.1}}%
      \put(6570,374){\makebox(0,0){\strut{} 1}}%
      \put(3796,1425){\rotatebox{90}{\makebox(0,0){\strut{}$F(t,t;p)$}}}%
      \put(5820,154){\makebox(0,0){\strut{}$p$}}%
      \put(5977,2090){\makebox(0,0)[l]{\strut{}$N=2$}}%
      \put(5977,1874){\makebox(0,0)[l]{\strut{}$\gamma=0.02$}}%
      \put(5977,1658){\makebox(0,0)[l]{\strut{}$t=1000$}}%
    }%
    \gplgaddtomacro\gplfronttext{%
      \csname LTb\endcsname%
      \put(5624,995){\makebox(0,0)[r]{\strut{}class.}}%
      \csname LTb\endcsname%
      \put(5624,775){\makebox(0,0)[r]{\strut{}texture}}%
    }%
    \gplbacktext
    \put(0,0){\includegraphics{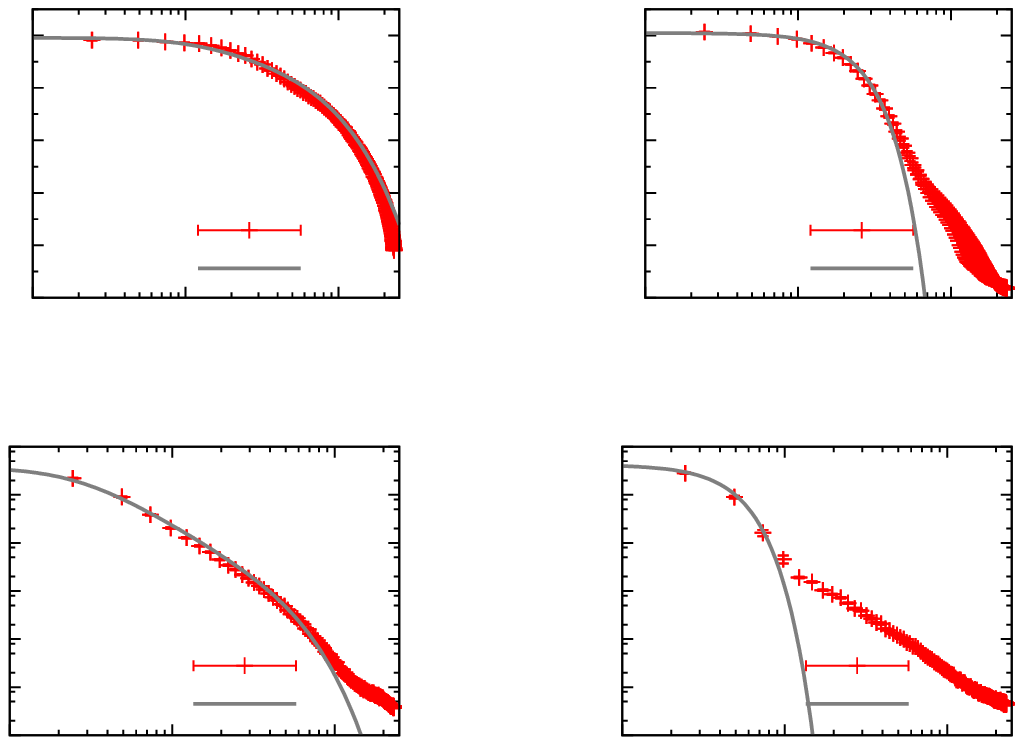}}%
    \gplfronttext
  \end{picture}%
\endgroup

%% file: new_field_final_N1_phi_log.tex
\begingroup
  \makeatletter
  \providecommand\color[2][]{%
    \GenericError{(gnuplot) \space\space\space\@spaces}{%
      Package color not loaded in conjunction with
      terminal option `colourtext'%
    }{See the gnuplot documentation for explanation.%
    }{Either use 'blacktext' in gnuplot or load the package
      color.sty in LaTeX.}%
    \renewcommand\color[2][]{}%
  }%
  \providecommand\includegraphics[2][]{%
    \GenericError{(gnuplot) \space\space\space\@spaces}{%
      Package graphicx or graphics not loaded%
    }{See the gnuplot documentation for explanation.%
    }{The gnuplot epslatex terminal needs graphicx.sty or graphics.sty.}%
    \renewcommand\includegraphics[2][]{}%
  }%
  \providecommand\rotatebox[2]{#2}%
  \@ifundefined{ifGPcolor}{%
    \newif\ifGPcolor
    \GPcolortrue
  }{}%
  \@ifundefined{ifGPblacktext}{%
    \newif\ifGPblacktext
    \GPblacktexttrue
  }{}%
  \let\gplgaddtomacro\g@addto@macro
  \gdef\gplbacktext{}%
  \gdef\gplfronttext{}%
  \makeatother
  \ifGPblacktext
    \def\colorrgb#1{}%
    \def\colorgray#1{}%
  \else
    \ifGPcolor
      \def\colorrgb#1{\color[rgb]{#1}}%
      \def\colorgray#1{\color[gray]{#1}}%
      \expandafter\def\csname LTw\endcsname{\color{white}}%
      \expandafter\def\csname LTb\endcsname{\color{black}}%
      \expandafter\def\csname LTa\endcsname{\color{black}}%
      \expandafter\def\csname LT0\endcsname{\color[rgb]{1,0,0}}%
      \expandafter\def\csname LT1\endcsname{\color[rgb]{0,1,0}}%
      \expandafter\def\csname LT2\endcsname{\color[rgb]{0,0,1}}%
      \expandafter\def\csname LT3\endcsname{\color[rgb]{1,0,1}}%
      \expandafter\def\csname LT4\endcsname{\color[rgb]{0,1,1}}%
      \expandafter\def\csname LT5\endcsname{\color[rgb]{1,1,0}}%
      \expandafter\def\csname LT6\endcsname{\color[rgb]{0,0,0}}%
      \expandafter\def\csname LT7\endcsname{\color[rgb]{1,0.3,0}}%
      \expandafter\def\csname LT8\endcsname{\color[rgb]{0.5,0.5,0.5}}%
    \else
      \def\colorrgb#1{\color{black}}%
      \def\colorgray#1{\color[gray]{#1}}%
      \expandafter\def\csname LTw\endcsname{\color{white}}%
      \expandafter\def\csname LTb\endcsname{\color{black}}%
      \expandafter\def\csname LTa\endcsname{\color{black}}%
      \expandafter\def\csname LT0\endcsname{\color{black}}%
      \expandafter\def\csname LT1\endcsname{\color{black}}%
      \expandafter\def\csname LT2\endcsname{\color{black}}%
      \expandafter\def\csname LT3\endcsname{\color{black}}%
      \expandafter\def\csname LT4\endcsname{\color{black}}%
      \expandafter\def\csname LT5\endcsname{\color{black}}%
      \expandafter\def\csname LT6\endcsname{\color{black}}%
      \expandafter\def\csname LT7\endcsname{\color{black}}%
      \expandafter\def\csname LT8\endcsname{\color{black}}%
    \fi
  \fi
  \setlength{\unitlength}{0.0500bp}%
  \begin{picture}(5760.00,3528.00)%
    \gplgaddtomacro\gplbacktext{%
      \csname LTb\endcsname%
      \put(1342,704){\makebox(0,0)[r]{\strut{} 0.01}}%
      \put(1342,1536){\makebox(0,0)[r]{\strut{} 0.1}}%
      \put(1342,2367){\makebox(0,0)[r]{\strut{} 1}}%
      \put(1474,484){\makebox(0,0){\strut{} 1}}%
      \put(3553,484){\makebox(0,0){\strut{} 10}}%
      \put(440,1786){\rotatebox{90}{\makebox(0,0){\strut{}$\phi(t)$}}}%
      \put(3452,154){\makebox(0,0){\strut{}$t$}}%
      \put(3452,3198){\makebox(0,0){\strut{}$N=1$}}%
    }%
    \gplgaddtomacro\gplfronttext{%
      \csname LTb\endcsname%
      \put(4443,1317){\makebox(0,0)[r]{\strut{}$J=0.01$}}%
      \csname LTb\endcsname%
      \put(4443,1097){\makebox(0,0)[r]{\strut{}$J=0.04$}}%
      \csname LTb\endcsname%
      \put(4443,877){\makebox(0,0)[r]{\strut{}$J=0.4$}}%
    }%
    \gplbacktext
    \put(0,0){\includegraphics{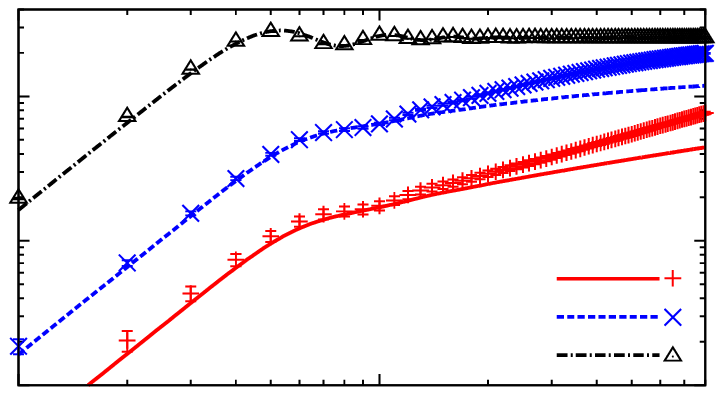}}%
    \gplfronttext
  \end{picture}%
\endgroup

%% file: new_field_final_N1_t10.tex
\begingroup
  \makeatletter
  \providecommand\color[2][]{%
    \GenericError{(gnuplot) \space\space\space\@spaces}{%
      Package color not loaded in conjunction with
      terminal option `colourtext'%
    }{See the gnuplot documentation for explanation.%
    }{Either use 'blacktext' in gnuplot or load the package
      color.sty in LaTeX.}%
    \renewcommand\color[2][]{}%
  }%
  \providecommand\includegraphics[2][]{%
    \GenericError{(gnuplot) \space\space\space\@spaces}{%
      Package graphicx or graphics not loaded%
    }{See the gnuplot documentation for explanation.%
    }{The gnuplot epslatex terminal needs graphicx.sty or graphics.sty.}%
    \renewcommand\includegraphics[2][]{}%
  }%
  \providecommand\rotatebox[2]{#2}%
  \@ifundefined{ifGPcolor}{%
    \newif\ifGPcolor
    \GPcolortrue
  }{}%
  \@ifundefined{ifGPblacktext}{%
    \newif\ifGPblacktext
    \GPblacktexttrue
  }{}%
  \let\gplgaddtomacro\g@addto@macro
  \gdef\gplbacktext{}%
  \gdef\gplfronttext{}%
  \makeatother
  \ifGPblacktext
    \def\colorrgb#1{}%
    \def\colorgray#1{}%
  \else
    \ifGPcolor
      \def\colorrgb#1{\color[rgb]{#1}}%
      \def\colorgray#1{\color[gray]{#1}}%
      \expandafter\def\csname LTw\endcsname{\color{white}}%
      \expandafter\def\csname LTb\endcsname{\color{black}}%
      \expandafter\def\csname LTa\endcsname{\color{black}}%
      \expandafter\def\csname LT0\endcsname{\color[rgb]{1,0,0}}%
      \expandafter\def\csname LT1\endcsname{\color[rgb]{0,1,0}}%
      \expandafter\def\csname LT2\endcsname{\color[rgb]{0,0,1}}%
      \expandafter\def\csname LT3\endcsname{\color[rgb]{1,0,1}}%
      \expandafter\def\csname LT4\endcsname{\color[rgb]{0,1,1}}%
      \expandafter\def\csname LT5\endcsname{\color[rgb]{1,1,0}}%
      \expandafter\def\csname LT6\endcsname{\color[rgb]{0,0,0}}%
      \expandafter\def\csname LT7\endcsname{\color[rgb]{1,0.3,0}}%
      \expandafter\def\csname LT8\endcsname{\color[rgb]{0.5,0.5,0.5}}%
    \else
      \def\colorrgb#1{\color{black}}%
      \def\colorgray#1{\color[gray]{#1}}%
      \expandafter\def\csname LTw\endcsname{\color{white}}%
      \expandafter\def\csname LTb\endcsname{\color{black}}%
      \expandafter\def\csname LTa\endcsname{\color{black}}%
      \expandafter\def\csname LT0\endcsname{\color{black}}%
      \expandafter\def\csname LT1\endcsname{\color{black}}%
      \expandafter\def\csname LT2\endcsname{\color{black}}%
      \expandafter\def\csname LT3\endcsname{\color{black}}%
      \expandafter\def\csname LT4\endcsname{\color{black}}%
      \expandafter\def\csname LT5\endcsname{\color{black}}%
      \expandafter\def\csname LT6\endcsname{\color{black}}%
      \expandafter\def\csname LT7\endcsname{\color{black}}%
      \expandafter\def\csname LT8\endcsname{\color{black}}%
    \fi
  \fi
  \setlength{\unitlength}{0.0500bp}%
  \begin{picture}(5760.00,3528.00)%
    \gplgaddtomacro\gplbacktext{%
      \csname LTb\endcsname%
      \put(1474,704){\makebox(0,0)[r]{\strut{} 0.001}}%
      \put(1474,1216){\makebox(0,0)[r]{\strut{} 0.01}}%
      \put(1474,1728){\makebox(0,0)[r]{\strut{} 0.1}}%
      \put(1474,2240){\makebox(0,0)[r]{\strut{} 1}}%
      \put(1474,2752){\makebox(0,0)[r]{\strut{} 10}}%
      \put(1474,3264){\makebox(0,0)[r]{\strut{} 100}}%
      \put(1606,484){\makebox(0,0){\strut{} 0.01}}%
      \put(3201,484){\makebox(0,0){\strut{} 0.1}}%
      \put(4795,484){\makebox(0,0){\strut{} 1}}%
      \put(440,1984){\rotatebox{90}{\makebox(0,0){\strut{}$F(t,t;p)$}}}%
      \put(3518,154){\makebox(0,0){\strut{}$p$}}%
      \put(3939,3059){\makebox(0,0)[l]{\strut{}$N=1, t=10$}}%
    }%
    \gplgaddtomacro\gplfronttext{%
      \csname LTb\endcsname%
      \put(2826,1350){\makebox(0,0)[r]{\strut{}$J=0$}}%
      \csname LTb\endcsname%
      \put(2826,1130){\makebox(0,0)[r]{\strut{}$J=0.04$}}%
      \csname LTb\endcsname%
      \put(2826,910){\makebox(0,0)[r]{\strut{}$J=0.4$}}%
    }%
    \gplbacktext
    \put(0,0){\includegraphics{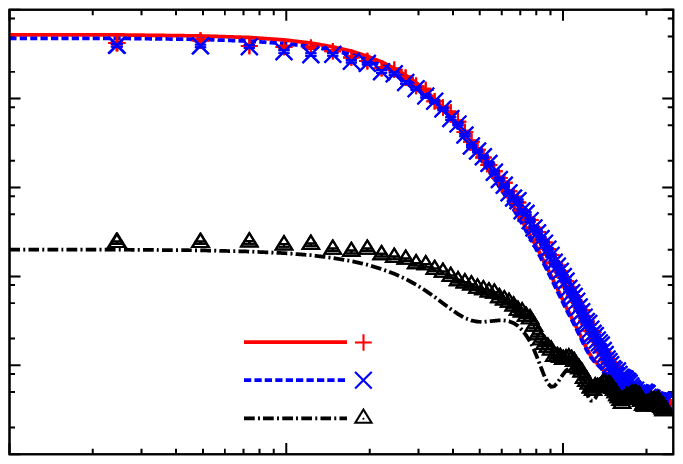}}%
    \gplfronttext
  \end{picture}%
\endgroup

%% file: new_field_final_N1_t40.tex
\begingroup
  \makeatletter
  \providecommand\color[2][]{%
    \GenericError{(gnuplot) \space\space\space\@spaces}{%
      Package color not loaded in conjunction with
      terminal option `colourtext'%
    }{See the gnuplot documentation for explanation.%
    }{Either use 'blacktext' in gnuplot or load the package
      color.sty in LaTeX.}%
    \renewcommand\color[2][]{}%
  }%
  \providecommand\includegraphics[2][]{%
    \GenericError{(gnuplot) \space\space\space\@spaces}{%
      Package graphicx or graphics not loaded%
    }{See the gnuplot documentation for explanation.%
    }{The gnuplot epslatex terminal needs graphicx.sty or graphics.sty.}%
    \renewcommand\includegraphics[2][]{}%
  }%
  \providecommand\rotatebox[2]{#2}%
  \@ifundefined{ifGPcolor}{%
    \newif\ifGPcolor
    \GPcolortrue
  }{}%
  \@ifundefined{ifGPblacktext}{%
    \newif\ifGPblacktext
    \GPblacktexttrue
  }{}%
  \let\gplgaddtomacro\g@addto@macro
  \gdef\gplbacktext{}%
  \gdef\gplfronttext{}%
  \makeatother
  \ifGPblacktext
    \def\colorrgb#1{}%
    \def\colorgray#1{}%
  \else
    \ifGPcolor
      \def\colorrgb#1{\color[rgb]{#1}}%
      \def\colorgray#1{\color[gray]{#1}}%
      \expandafter\def\csname LTw\endcsname{\color{white}}%
      \expandafter\def\csname LTb\endcsname{\color{black}}%
      \expandafter\def\csname LTa\endcsname{\color{black}}%
      \expandafter\def\csname LT0\endcsname{\color[rgb]{1,0,0}}%
      \expandafter\def\csname LT1\endcsname{\color[rgb]{0,1,0}}%
      \expandafter\def\csname LT2\endcsname{\color[rgb]{0,0,1}}%
      \expandafter\def\csname LT3\endcsname{\color[rgb]{1,0,1}}%
      \expandafter\def\csname LT4\endcsname{\color[rgb]{0,1,1}}%
      \expandafter\def\csname LT5\endcsname{\color[rgb]{1,1,0}}%
      \expandafter\def\csname LT6\endcsname{\color[rgb]{0,0,0}}%
      \expandafter\def\csname LT7\endcsname{\color[rgb]{1,0.3,0}}%
      \expandafter\def\csname LT8\endcsname{\color[rgb]{0.5,0.5,0.5}}%
    \else
      \def\colorrgb#1{\color{black}}%
      \def\colorgray#1{\color[gray]{#1}}%
      \expandafter\def\csname LTw\endcsname{\color{white}}%
      \expandafter\def\csname LTb\endcsname{\color{black}}%
      \expandafter\def\csname LTa\endcsname{\color{black}}%
      \expandafter\def\csname LT0\endcsname{\color{black}}%
      \expandafter\def\csname LT1\endcsname{\color{black}}%
      \expandafter\def\csname LT2\endcsname{\color{black}}%
      \expandafter\def\csname LT3\endcsname{\color{black}}%
      \expandafter\def\csname LT4\endcsname{\color{black}}%
      \expandafter\def\csname LT5\endcsname{\color{black}}%
      \expandafter\def\csname LT6\endcsname{\color{black}}%
      \expandafter\def\csname LT7\endcsname{\color{black}}%
      \expandafter\def\csname LT8\endcsname{\color{black}}%
    \fi
  \fi
  \setlength{\unitlength}{0.0500bp}%
  \begin{picture}(5760.00,3528.00)%
    \gplgaddtomacro\gplbacktext{%
      \csname LTb\endcsname%
      \put(1606,704){\makebox(0,0)[r]{\strut{} 1e-08}}%
      \put(1606,1169){\makebox(0,0)[r]{\strut{} 1e-06}}%
      \put(1606,1635){\makebox(0,0)[r]{\strut{} 0.0001}}%
      \put(1606,2100){\makebox(0,0)[r]{\strut{} 0.01}}%
      \put(1606,2566){\makebox(0,0)[r]{\strut{} 1}}%
      \put(1606,3031){\makebox(0,0)[r]{\strut{} 100}}%
      \put(1738,484){\makebox(0,0){\strut{} 0.01}}%
      \put(3278,484){\makebox(0,0){\strut{} 0.1}}%
      \put(4817,484){\makebox(0,0){\strut{} 1}}%
      \put(440,1984){\rotatebox{90}{\makebox(0,0){\strut{}$F(t,t;p)$}}}%
      \put(3584,154){\makebox(0,0){\strut{}$p$}}%
      \put(3990,3059){\makebox(0,0)[l]{\strut{}$N=1, t=40$}}%
    }%
    \gplgaddtomacro\gplfronttext{%
      \csname LTb\endcsname%
      \put(2886,2153){\makebox(0,0)[r]{\strut{}$J=0$}}%
      \csname LTb\endcsname%
      \put(2886,1933){\makebox(0,0)[r]{\strut{}$J=0.04$}}%
      \csname LTb\endcsname%
      \put(2886,1713){\makebox(0,0)[r]{\strut{}$J=0.4$}}%
    }%
    \gplbacktext
    \put(0,0){\includegraphics{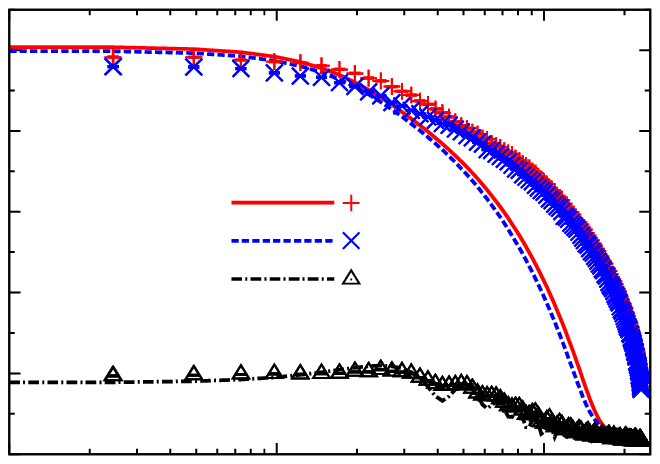}}%
    \gplfronttext
  \end{picture}%
\endgroup

%% file: new_field_final_N2_phi_log.tex
\begingroup
  \makeatletter
  \providecommand\color[2][]{%
    \GenericError{(gnuplot) \space\space\space\@spaces}{%
      Package color not loaded in conjunction with
      terminal option `colourtext'%
    }{See the gnuplot documentation for explanation.%
    }{Either use 'blacktext' in gnuplot or load the package
      color.sty in LaTeX.}%
    \renewcommand\color[2][]{}%
  }%
  \providecommand\includegraphics[2][]{%
    \GenericError{(gnuplot) \space\space\space\@spaces}{%
      Package graphicx or graphics not loaded%
    }{See the gnuplot documentation for explanation.%
    }{The gnuplot epslatex terminal needs graphicx.sty or graphics.sty.}%
    \renewcommand\includegraphics[2][]{}%
  }%
  \providecommand\rotatebox[2]{#2}%
  \@ifundefined{ifGPcolor}{%
    \newif\ifGPcolor
    \GPcolortrue
  }{}%
  \@ifundefined{ifGPblacktext}{%
    \newif\ifGPblacktext
    \GPblacktexttrue
  }{}%
  \let\gplgaddtomacro\g@addto@macro
  \gdef\gplbacktext{}%
  \gdef\gplfronttext{}%
  \makeatother
  \ifGPblacktext
    \def\colorrgb#1{}%
    \def\colorgray#1{}%
  \else
    \ifGPcolor
      \def\colorrgb#1{\color[rgb]{#1}}%
      \def\colorgray#1{\color[gray]{#1}}%
      \expandafter\def\csname LTw\endcsname{\color{white}}%
      \expandafter\def\csname LTb\endcsname{\color{black}}%
      \expandafter\def\csname LTa\endcsname{\color{black}}%
      \expandafter\def\csname LT0\endcsname{\color[rgb]{1,0,0}}%
      \expandafter\def\csname LT1\endcsname{\color[rgb]{0,1,0}}%
      \expandafter\def\csname LT2\endcsname{\color[rgb]{0,0,1}}%
      \expandafter\def\csname LT3\endcsname{\color[rgb]{1,0,1}}%
      \expandafter\def\csname LT4\endcsname{\color[rgb]{0,1,1}}%
      \expandafter\def\csname LT5\endcsname{\color[rgb]{1,1,0}}%
      \expandafter\def\csname LT6\endcsname{\color[rgb]{0,0,0}}%
      \expandafter\def\csname LT7\endcsname{\color[rgb]{1,0.3,0}}%
      \expandafter\def\csname LT8\endcsname{\color[rgb]{0.5,0.5,0.5}}%
    \else
      \def\colorrgb#1{\color{black}}%
      \def\colorgray#1{\color[gray]{#1}}%
      \expandafter\def\csname LTw\endcsname{\color{white}}%
      \expandafter\def\csname LTb\endcsname{\color{black}}%
      \expandafter\def\csname LTa\endcsname{\color{black}}%
      \expandafter\def\csname LT0\endcsname{\color{black}}%
      \expandafter\def\csname LT1\endcsname{\color{black}}%
      \expandafter\def\csname LT2\endcsname{\color{black}}%
      \expandafter\def\csname LT3\endcsname{\color{black}}%
      \expandafter\def\csname LT4\endcsname{\color{black}}%
      \expandafter\def\csname LT5\endcsname{\color{black}}%
      \expandafter\def\csname LT6\endcsname{\color{black}}%
      \expandafter\def\csname LT7\endcsname{\color{black}}%
      \expandafter\def\csname LT8\endcsname{\color{black}}%
    \fi
  \fi
  \setlength{\unitlength}{0.0500bp}%
  \begin{picture}(5760.00,3528.00)%
    \gplgaddtomacro\gplbacktext{%
      \csname LTb\endcsname%
      \put(1342,704){\makebox(0,0)[r]{\strut{} 0.01}}%
      \put(1342,1506){\makebox(0,0)[r]{\strut{} 0.1}}%
      \put(1342,2308){\makebox(0,0)[r]{\strut{} 1}}%
      \put(1474,484){\makebox(0,0){\strut{} 1}}%
      \put(3618,484){\makebox(0,0){\strut{} 10}}%
      \put(440,1786){\rotatebox{90}{\makebox(0,0){\strut{}$\phi(t)$}}}%
      \put(3452,154){\makebox(0,0){\strut{}$t$}}%
      \put(3452,3198){\makebox(0,0){\strut{}$N=2$}}%
    }%
    \gplgaddtomacro\gplfronttext{%
      \csname LTb\endcsname%
      \put(4443,1317){\makebox(0,0)[r]{\strut{}$J=0.01$}}%
      \csname LTb\endcsname%
      \put(4443,1097){\makebox(0,0)[r]{\strut{}$J=0.04$}}%
      \csname LTb\endcsname%
      \put(4443,877){\makebox(0,0)[r]{\strut{}$J=0.4$}}%
    }%
    \gplbacktext
    \put(0,0){\includegraphics{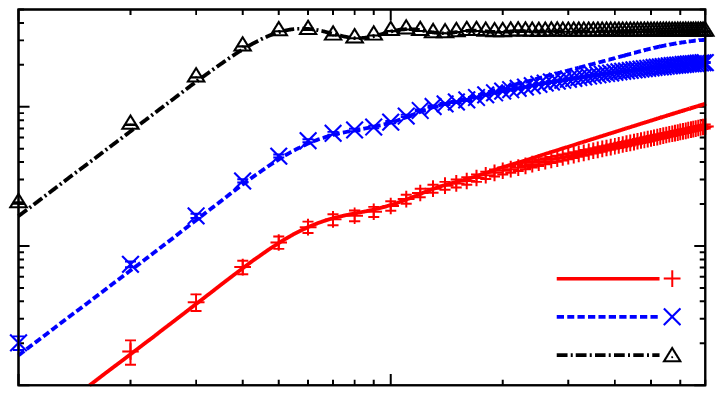}}%
    \gplfronttext
  \end{picture}%
\endgroup

%% file: new_field_final_N2_t10.tex
\begingroup
  \makeatletter
  \providecommand\color[2][]{%
    \GenericError{(gnuplot) \space\space\space\@spaces}{%
      Package color not loaded in conjunction with
      terminal option `colourtext'%
    }{See the gnuplot documentation for explanation.%
    }{Either use 'blacktext' in gnuplot or load the package
      color.sty in LaTeX.}%
    \renewcommand\color[2][]{}%
  }%
  \providecommand\includegraphics[2][]{%
    \GenericError{(gnuplot) \space\space\space\@spaces}{%
      Package graphicx or graphics not loaded%
    }{See the gnuplot documentation for explanation.%
    }{The gnuplot epslatex terminal needs graphicx.sty or graphics.sty.}%
    \renewcommand\includegraphics[2][]{}%
  }%
  \providecommand\rotatebox[2]{#2}%
  \@ifundefined{ifGPcolor}{%
    \newif\ifGPcolor
    \GPcolortrue
  }{}%
  \@ifundefined{ifGPblacktext}{%
    \newif\ifGPblacktext
    \GPblacktexttrue
  }{}%
  \let\gplgaddtomacro\g@addto@macro
  \gdef\gplbacktext{}%
  \gdef\gplfronttext{}%
  \makeatother
  \ifGPblacktext
    \def\colorrgb#1{}%
    \def\colorgray#1{}%
  \else
    \ifGPcolor
      \def\colorrgb#1{\color[rgb]{#1}}%
      \def\colorgray#1{\color[gray]{#1}}%
      \expandafter\def\csname LTw\endcsname{\color{white}}%
      \expandafter\def\csname LTb\endcsname{\color{black}}%
      \expandafter\def\csname LTa\endcsname{\color{black}}%
      \expandafter\def\csname LT0\endcsname{\color[rgb]{1,0,0}}%
      \expandafter\def\csname LT1\endcsname{\color[rgb]{0,1,0}}%
      \expandafter\def\csname LT2\endcsname{\color[rgb]{0,0,1}}%
      \expandafter\def\csname LT3\endcsname{\color[rgb]{1,0,1}}%
      \expandafter\def\csname LT4\endcsname{\color[rgb]{0,1,1}}%
      \expandafter\def\csname LT5\endcsname{\color[rgb]{1,1,0}}%
      \expandafter\def\csname LT6\endcsname{\color[rgb]{0,0,0}}%
      \expandafter\def\csname LT7\endcsname{\color[rgb]{1,0.3,0}}%
      \expandafter\def\csname LT8\endcsname{\color[rgb]{0.5,0.5,0.5}}%
    \else
      \def\colorrgb#1{\color{black}}%
      \def\colorgray#1{\color[gray]{#1}}%
      \expandafter\def\csname LTw\endcsname{\color{white}}%
      \expandafter\def\csname LTb\endcsname{\color{black}}%
      \expandafter\def\csname LTa\endcsname{\color{black}}%
      \expandafter\def\csname LT0\endcsname{\color{black}}%
      \expandafter\def\csname LT1\endcsname{\color{black}}%
      \expandafter\def\csname LT2\endcsname{\color{black}}%
      \expandafter\def\csname LT3\endcsname{\color{black}}%
      \expandafter\def\csname LT4\endcsname{\color{black}}%
      \expandafter\def\csname LT5\endcsname{\color{black}}%
      \expandafter\def\csname LT6\endcsname{\color{black}}%
      \expandafter\def\csname LT7\endcsname{\color{black}}%
      \expandafter\def\csname LT8\endcsname{\color{black}}%
    \fi
  \fi
  \setlength{\unitlength}{0.0500bp}%
  \begin{picture}(5760.00,3528.00)%
    \gplgaddtomacro\gplbacktext{%
      \csname LTb\endcsname%
      \put(1474,704){\makebox(0,0)[r]{\strut{} 0.001}}%
      \put(1474,1216){\makebox(0,0)[r]{\strut{} 0.01}}%
      \put(1474,1728){\makebox(0,0)[r]{\strut{} 0.1}}%
      \put(1474,2240){\makebox(0,0)[r]{\strut{} 1}}%
      \put(1474,2752){\makebox(0,0)[r]{\strut{} 10}}%
      \put(1474,3264){\makebox(0,0)[r]{\strut{} 100}}%
      \put(1606,484){\makebox(0,0){\strut{} 0.01}}%
      \put(3201,484){\makebox(0,0){\strut{} 0.1}}%
      \put(4795,484){\makebox(0,0){\strut{} 1}}%
      \put(440,1984){\rotatebox{90}{\makebox(0,0){\strut{}$F_{\parallel}(t,t;p)$}}}%
      \put(3518,154){\makebox(0,0){\strut{}$p$}}%
      \put(3939,3059){\makebox(0,0)[l]{\strut{}$N=2, t=10$}}%
    }%
    \gplgaddtomacro\gplfronttext{%
      \csname LTb\endcsname%
      \put(2826,1350){\makebox(0,0)[r]{\strut{}$J=0$}}%
      \csname LTb\endcsname%
      \put(2826,1130){\makebox(0,0)[r]{\strut{}$J=0.04$}}%
      \csname LTb\endcsname%
      \put(2826,910){\makebox(0,0)[r]{\strut{}$J=0.4$}}%
    }%
    \gplbacktext
    \put(0,0){\includegraphics{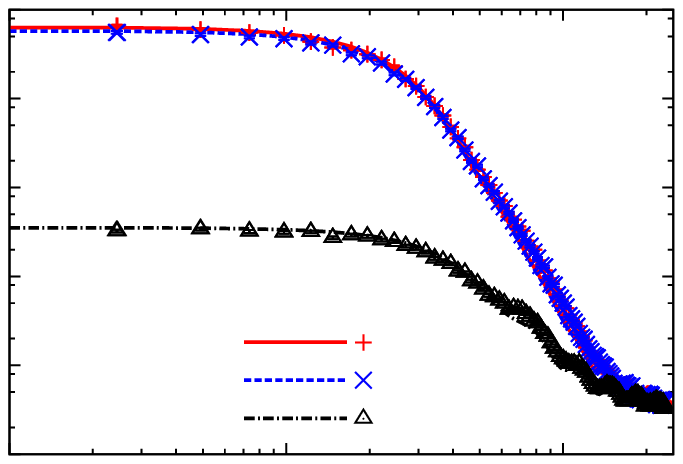}}%
    \gplfronttext
  \end{picture}%
\endgroup

%% file: new_field_final_N2_t40.tex
\begingroup
  \makeatletter
  \providecommand\color[2][]{%
    \GenericError{(gnuplot) \space\space\space\@spaces}{%
      Package color not loaded in conjunction with
      terminal option `colourtext'%
    }{See the gnuplot documentation for explanation.%
    }{Either use 'blacktext' in gnuplot or load the package
      color.sty in LaTeX.}%
    \renewcommand\color[2][]{}%
  }%
  \providecommand\includegraphics[2][]{%
    \GenericError{(gnuplot) \space\space\space\@spaces}{%
      Package graphicx or graphics not loaded%
    }{See the gnuplot documentation for explanation.%
    }{The gnuplot epslatex terminal needs graphicx.sty or graphics.sty.}%
    \renewcommand\includegraphics[2][]{}%
  }%
  \providecommand\rotatebox[2]{#2}%
  \@ifundefined{ifGPcolor}{%
    \newif\ifGPcolor
    \GPcolortrue
  }{}%
  \@ifundefined{ifGPblacktext}{%
    \newif\ifGPblacktext
    \GPblacktexttrue
  }{}%
  \let\gplgaddtomacro\g@addto@macro
  \gdef\gplbacktext{}%
  \gdef\gplfronttext{}%
  \makeatother
  \ifGPblacktext
    \def\colorrgb#1{}%
    \def\colorgray#1{}%
  \else
    \ifGPcolor
      \def\colorrgb#1{\color[rgb]{#1}}%
      \def\colorgray#1{\color[gray]{#1}}%
      \expandafter\def\csname LTw\endcsname{\color{white}}%
      \expandafter\def\csname LTb\endcsname{\color{black}}%
      \expandafter\def\csname LTa\endcsname{\color{black}}%
      \expandafter\def\csname LT0\endcsname{\color[rgb]{1,0,0}}%
      \expandafter\def\csname LT1\endcsname{\color[rgb]{0,1,0}}%
      \expandafter\def\csname LT2\endcsname{\color[rgb]{0,0,1}}%
      \expandafter\def\csname LT3\endcsname{\color[rgb]{1,0,1}}%
      \expandafter\def\csname LT4\endcsname{\color[rgb]{0,1,1}}%
      \expandafter\def\csname LT5\endcsname{\color[rgb]{1,1,0}}%
      \expandafter\def\csname LT6\endcsname{\color[rgb]{0,0,0}}%
      \expandafter\def\csname LT7\endcsname{\color[rgb]{1,0.3,0}}%
      \expandafter\def\csname LT8\endcsname{\color[rgb]{0.5,0.5,0.5}}%
    \else
      \def\colorrgb#1{\color{black}}%
      \def\colorgray#1{\color[gray]{#1}}%
      \expandafter\def\csname LTw\endcsname{\color{white}}%
      \expandafter\def\csname LTb\endcsname{\color{black}}%
      \expandafter\def\csname LTa\endcsname{\color{black}}%
      \expandafter\def\csname LT0\endcsname{\color{black}}%
      \expandafter\def\csname LT1\endcsname{\color{black}}%
      \expandafter\def\csname LT2\endcsname{\color{black}}%
      \expandafter\def\csname LT3\endcsname{\color{black}}%
      \expandafter\def\csname LT4\endcsname{\color{black}}%
      \expandafter\def\csname LT5\endcsname{\color{black}}%
      \expandafter\def\csname LT6\endcsname{\color{black}}%
      \expandafter\def\csname LT7\endcsname{\color{black}}%
      \expandafter\def\csname LT8\endcsname{\color{black}}%
    \fi
  \fi
  \setlength{\unitlength}{0.0500bp}%
  \begin{picture}(5760.00,3528.00)%
    \gplgaddtomacro\gplbacktext{%
      \csname LTb\endcsname%
      \put(1606,704){\makebox(0,0)[r]{\strut{} 1e-08}}%
      \put(1606,1169){\makebox(0,0)[r]{\strut{} 1e-06}}%
      \put(1606,1635){\makebox(0,0)[r]{\strut{} 0.0001}}%
      \put(1606,2100){\makebox(0,0)[r]{\strut{} 0.01}}%
      \put(1606,2566){\makebox(0,0)[r]{\strut{} 1}}%
      \put(1606,3031){\makebox(0,0)[r]{\strut{} 100}}%
      \put(1738,484){\makebox(0,0){\strut{} 0.01}}%
      \put(3278,484){\makebox(0,0){\strut{} 0.1}}%
      \put(4817,484){\makebox(0,0){\strut{} 1}}%
      \put(440,1984){\rotatebox{90}{\makebox(0,0){\strut{}$F_{\parallel}(t,t;p)$}}}%
      \put(3584,154){\makebox(0,0){\strut{}$p$}}%
      \put(3990,3059){\makebox(0,0)[l]{\strut{}$N=2, t=40$}}%
    }%
    \gplgaddtomacro\gplfronttext{%
      \csname LTb\endcsname%
      \put(2886,2153){\makebox(0,0)[r]{\strut{}$J=0$}}%
      \csname LTb\endcsname%
      \put(2886,1933){\makebox(0,0)[r]{\strut{}$J=0.04$}}%
      \csname LTb\endcsname%
      \put(2886,1713){\makebox(0,0)[r]{\strut{}$J=0.4$}}%
    }%
    \gplbacktext
    \put(0,0){\includegraphics{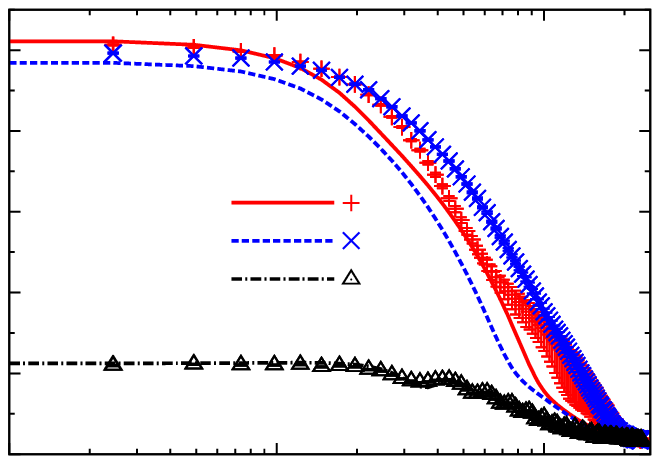}}%
    \gplfronttext
  \end{picture}%
\endgroup

%% file: new_field_final_N2_t10_trans.tex
\begingroup
  \makeatletter
  \providecommand\color[2][]{%
    \GenericError{(gnuplot) \space\space\space\@spaces}{%
      Package color not loaded in conjunction with
      terminal option `colourtext'%
    }{See the gnuplot documentation for explanation.%
    }{Either use 'blacktext' in gnuplot or load the package
      color.sty in LaTeX.}%
    \renewcommand\color[2][]{}%
  }%
  \providecommand\includegraphics[2][]{%
    \GenericError{(gnuplot) \space\space\space\@spaces}{%
      Package graphicx or graphics not loaded%
    }{See the gnuplot documentation for explanation.%
    }{The gnuplot epslatex terminal needs graphicx.sty or graphics.sty.}%
    \renewcommand\includegraphics[2][]{}%
  }%
  \providecommand\rotatebox[2]{#2}%
  \@ifundefined{ifGPcolor}{%
    \newif\ifGPcolor
    \GPcolortrue
  }{}%
  \@ifundefined{ifGPblacktext}{%
    \newif\ifGPblacktext
    \GPblacktexttrue
  }{}%
  \let\gplgaddtomacro\g@addto@macro
  \gdef\gplbacktext{}%
  \gdef\gplfronttext{}%
  \makeatother
  \ifGPblacktext
    \def\colorrgb#1{}%
    \def\colorgray#1{}%
  \else
    \ifGPcolor
      \def\colorrgb#1{\color[rgb]{#1}}%
      \def\colorgray#1{\color[gray]{#1}}%
      \expandafter\def\csname LTw\endcsname{\color{white}}%
      \expandafter\def\csname LTb\endcsname{\color{black}}%
      \expandafter\def\csname LTa\endcsname{\color{black}}%
      \expandafter\def\csname LT0\endcsname{\color[rgb]{1,0,0}}%
      \expandafter\def\csname LT1\endcsname{\color[rgb]{0,1,0}}%
      \expandafter\def\csname LT2\endcsname{\color[rgb]{0,0,1}}%
      \expandafter\def\csname LT3\endcsname{\color[rgb]{1,0,1}}%
      \expandafter\def\csname LT4\endcsname{\color[rgb]{0,1,1}}%
      \expandafter\def\csname LT5\endcsname{\color[rgb]{1,1,0}}%
      \expandafter\def\csname LT6\endcsname{\color[rgb]{0,0,0}}%
      \expandafter\def\csname LT7\endcsname{\color[rgb]{1,0.3,0}}%
      \expandafter\def\csname LT8\endcsname{\color[rgb]{0.5,0.5,0.5}}%
    \else
      \def\colorrgb#1{\color{black}}%
      \def\colorgray#1{\color[gray]{#1}}%
      \expandafter\def\csname LTw\endcsname{\color{white}}%
      \expandafter\def\csname LTb\endcsname{\color{black}}%
      \expandafter\def\csname LTa\endcsname{\color{black}}%
      \expandafter\def\csname LT0\endcsname{\color{black}}%
      \expandafter\def\csname LT1\endcsname{\color{black}}%
      \expandafter\def\csname LT2\endcsname{\color{black}}%
      \expandafter\def\csname LT3\endcsname{\color{black}}%
      \expandafter\def\csname LT4\endcsname{\color{black}}%
      \expandafter\def\csname LT5\endcsname{\color{black}}%
      \expandafter\def\csname LT6\endcsname{\color{black}}%
      \expandafter\def\csname LT7\endcsname{\color{black}}%
      \expandafter\def\csname LT8\endcsname{\color{black}}%
    \fi
  \fi
  \setlength{\unitlength}{0.0500bp}%
  \begin{picture}(5760.00,3528.00)%
    \gplgaddtomacro\gplbacktext{%
      \csname LTb\endcsname%
      \put(1474,704){\makebox(0,0)[r]{\strut{} 0.001}}%
      \put(1474,1216){\makebox(0,0)[r]{\strut{} 0.01}}%
      \put(1474,1728){\makebox(0,0)[r]{\strut{} 0.1}}%
      \put(1474,2240){\makebox(0,0)[r]{\strut{} 1}}%
      \put(1474,2752){\makebox(0,0)[r]{\strut{} 10}}%
      \put(1474,3264){\makebox(0,0)[r]{\strut{} 100}}%
      \put(1606,484){\makebox(0,0){\strut{} 0.01}}%
      \put(3201,484){\makebox(0,0){\strut{} 0.1}}%
      \put(4795,484){\makebox(0,0){\strut{} 1}}%
      \put(440,1984){\rotatebox{90}{\makebox(0,0){\strut{}$F_{\perp}(t,t;p)$}}}%
      \put(3518,154){\makebox(0,0){\strut{}$p$}}%
      \put(3939,3059){\makebox(0,0)[l]{\strut{}$N=2, t=10$}}%
    }%
    \gplgaddtomacro\gplfronttext{%
      \csname LTb\endcsname%
      \put(2826,1350){\makebox(0,0)[r]{\strut{}$J=0$}}%
      \csname LTb\endcsname%
      \put(2826,1130){\makebox(0,0)[r]{\strut{}$J=0.04$}}%
      \csname LTb\endcsname%
      \put(2826,910){\makebox(0,0)[r]{\strut{}$J=0.4$}}%
    }%
    \gplbacktext
    \put(0,0){\includegraphics{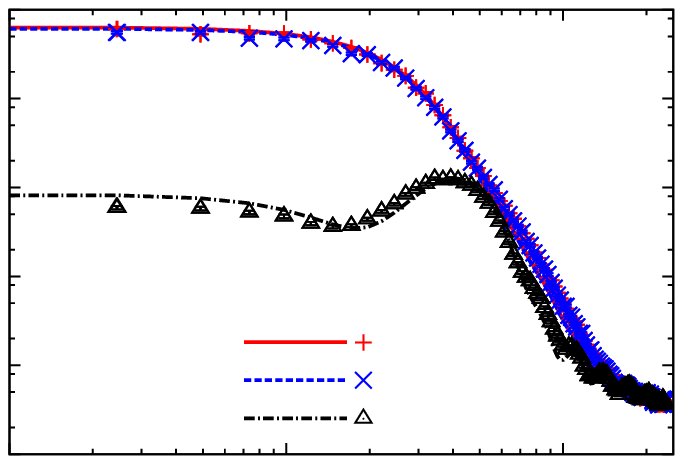}}%
    \gplfronttext
  \end{picture}%
\endgroup

%% file: new_field_final_N2_t40_trans.tex
\begingroup
  \makeatletter
  \providecommand\color[2][]{%
    \GenericError{(gnuplot) \space\space\space\@spaces}{%
      Package color not loaded in conjunction with
      terminal option `colourtext'%
    }{See the gnuplot documentation for explanation.%
    }{Either use 'blacktext' in gnuplot or load the package
      color.sty in LaTeX.}%
    \renewcommand\color[2][]{}%
  }%
  \providecommand\includegraphics[2][]{%
    \GenericError{(gnuplot) \space\space\space\@spaces}{%
      Package graphicx or graphics not loaded%
    }{See the gnuplot documentation for explanation.%
    }{The gnuplot epslatex terminal needs graphicx.sty or graphics.sty.}%
    \renewcommand\includegraphics[2][]{}%
  }%
  \providecommand\rotatebox[2]{#2}%
  \@ifundefined{ifGPcolor}{%
    \newif\ifGPcolor
    \GPcolortrue
  }{}%
  \@ifundefined{ifGPblacktext}{%
    \newif\ifGPblacktext
    \GPblacktexttrue
  }{}%
  \let\gplgaddtomacro\g@addto@macro
  \gdef\gplbacktext{}%
  \gdef\gplfronttext{}%
  \makeatother
  \ifGPblacktext
    \def\colorrgb#1{}%
    \def\colorgray#1{}%
  \else
    \ifGPcolor
      \def\colorrgb#1{\color[rgb]{#1}}%
      \def\colorgray#1{\color[gray]{#1}}%
      \expandafter\def\csname LTw\endcsname{\color{white}}%
      \expandafter\def\csname LTb\endcsname{\color{black}}%
      \expandafter\def\csname LTa\endcsname{\color{black}}%
      \expandafter\def\csname LT0\endcsname{\color[rgb]{1,0,0}}%
      \expandafter\def\csname LT1\endcsname{\color[rgb]{0,1,0}}%
      \expandafter\def\csname LT2\endcsname{\color[rgb]{0,0,1}}%
      \expandafter\def\csname LT3\endcsname{\color[rgb]{1,0,1}}%
      \expandafter\def\csname LT4\endcsname{\color[rgb]{0,1,1}}%
      \expandafter\def\csname LT5\endcsname{\color[rgb]{1,1,0}}%
      \expandafter\def\csname LT6\endcsname{\color[rgb]{0,0,0}}%
      \expandafter\def\csname LT7\endcsname{\color[rgb]{1,0.3,0}}%
      \expandafter\def\csname LT8\endcsname{\color[rgb]{0.5,0.5,0.5}}%
    \else
      \def\colorrgb#1{\color{black}}%
      \def\colorgray#1{\color[gray]{#1}}%
      \expandafter\def\csname LTw\endcsname{\color{white}}%
      \expandafter\def\csname LTb\endcsname{\color{black}}%
      \expandafter\def\csname LTa\endcsname{\color{black}}%
      \expandafter\def\csname LT0\endcsname{\color{black}}%
      \expandafter\def\csname LT1\endcsname{\color{black}}%
      \expandafter\def\csname LT2\endcsname{\color{black}}%
      \expandafter\def\csname LT3\endcsname{\color{black}}%
      \expandafter\def\csname LT4\endcsname{\color{black}}%
      \expandafter\def\csname LT5\endcsname{\color{black}}%
      \expandafter\def\csname LT6\endcsname{\color{black}}%
      \expandafter\def\csname LT7\endcsname{\color{black}}%
      \expandafter\def\csname LT8\endcsname{\color{black}}%
    \fi
  \fi
  \setlength{\unitlength}{0.0500bp}%
  \begin{picture}(5760.00,3528.00)%
    \gplgaddtomacro\gplbacktext{%
      \csname LTb\endcsname%
      \put(1606,704){\makebox(0,0)[r]{\strut{} 1e-08}}%
      \put(1606,1169){\makebox(0,0)[r]{\strut{} 1e-06}}%
      \put(1606,1635){\makebox(0,0)[r]{\strut{} 0.0001}}%
      \put(1606,2100){\makebox(0,0)[r]{\strut{} 0.01}}%
      \put(1606,2566){\makebox(0,0)[r]{\strut{} 1}}%
      \put(1606,3031){\makebox(0,0)[r]{\strut{} 100}}%
      \put(1738,484){\makebox(0,0){\strut{} 0.01}}%
      \put(3278,484){\makebox(0,0){\strut{} 0.1}}%
      \put(4817,484){\makebox(0,0){\strut{} 1}}%
      \put(440,1984){\rotatebox{90}{\makebox(0,0){\strut{}$F_{\perp}(t,t;p)$}}}%
      \put(3584,154){\makebox(0,0){\strut{}$p$}}%
      \put(3990,3059){\makebox(0,0)[l]{\strut{}$N=2, t=40$}}%
    }%
    \gplgaddtomacro\gplfronttext{%
      \csname LTb\endcsname%
      \put(2886,2153){\makebox(0,0)[r]{\strut{}$J=0$}}%
      \csname LTb\endcsname%
      \put(2886,1933){\makebox(0,0)[r]{\strut{}$J=0.04$}}%
      \csname LTb\endcsname%
      \put(2886,1713){\makebox(0,0)[r]{\strut{}$J=0.4$}}%
    }%
    \gplbacktext
    \put(0,0){\includegraphics{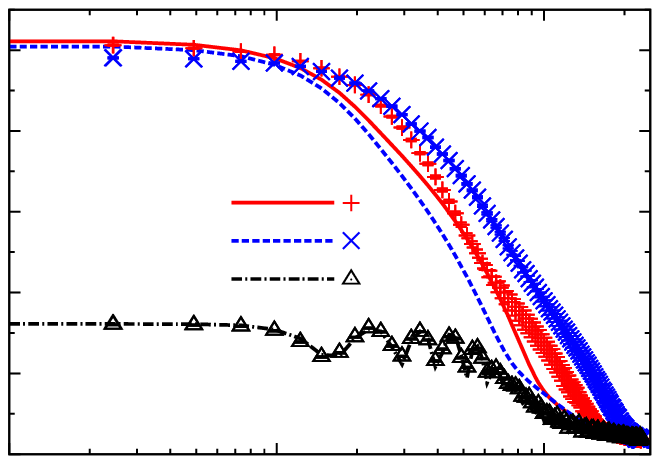}}%
    \gplfronttext
  \end{picture}%
\endgroup

%% file: inhom.tex
\begingroup
  \makeatletter
  \providecommand\color[2][]{%
    \GenericError{(gnuplot) \space\space\space\@spaces}{%
      Package color not loaded in conjunction with
      terminal option `colourtext'%
    }{See the gnuplot documentation for explanation.%
    }{Either use 'blacktext' in gnuplot or load the package
      color.sty in LaTeX.}%
    \renewcommand\color[2][]{}%
  }%
  \providecommand\includegraphics[2][]{%
    \GenericError{(gnuplot) \space\space\space\@spaces}{%
      Package graphicx or graphics not loaded%
    }{See the gnuplot documentation for explanation.%
    }{The gnuplot epslatex terminal needs graphicx.sty or graphics.sty.}%
    \renewcommand\includegraphics[2][]{}%
  }%
  \providecommand\rotatebox[2]{#2}%
  \@ifundefined{ifGPcolor}{%
    \newif\ifGPcolor
    \GPcolortrue
  }{}%
  \@ifundefined{ifGPblacktext}{%
    \newif\ifGPblacktext
    \GPblacktexttrue
  }{}%
  \let\gplgaddtomacro\g@addto@macro
  \gdef\gplbacktext{}%
  \gdef\gplfronttext{}%
  \makeatother
  \ifGPblacktext
    \def\colorrgb#1{}%
    \def\colorgray#1{}%
  \else
    \ifGPcolor
      \def\colorrgb#1{\color[rgb]{#1}}%
      \def\colorgray#1{\color[gray]{#1}}%
      \expandafter\def\csname LTw\endcsname{\color{white}}%
      \expandafter\def\csname LTb\endcsname{\color{black}}%
      \expandafter\def\csname LTa\endcsname{\color{black}}%
      \expandafter\def\csname LT0\endcsname{\color[rgb]{1,0,0}}%
      \expandafter\def\csname LT1\endcsname{\color[rgb]{0,1,0}}%
      \expandafter\def\csname LT2\endcsname{\color[rgb]{0,0,1}}%
      \expandafter\def\csname LT3\endcsname{\color[rgb]{1,0,1}}%
      \expandafter\def\csname LT4\endcsname{\color[rgb]{0,1,1}}%
      \expandafter\def\csname LT5\endcsname{\color[rgb]{1,1,0}}%
      \expandafter\def\csname LT6\endcsname{\color[rgb]{0,0,0}}%
      \expandafter\def\csname LT7\endcsname{\color[rgb]{1,0.3,0}}%
      \expandafter\def\csname LT8\endcsname{\color[rgb]{0.5,0.5,0.5}}%
    \else
      \def\colorrgb#1{\color{black}}%
      \def\colorgray#1{\color[gray]{#1}}%
      \expandafter\def\csname LTw\endcsname{\color{white}}%
      \expandafter\def\csname LTb\endcsname{\color{black}}%
      \expandafter\def\csname LTa\endcsname{\color{black}}%
      \expandafter\def\csname LT0\endcsname{\color{black}}%
      \expandafter\def\csname LT1\endcsname{\color{black}}%
      \expandafter\def\csname LT2\endcsname{\color{black}}%
      \expandafter\def\csname LT3\endcsname{\color{black}}%
      \expandafter\def\csname LT4\endcsname{\color{black}}%
      \expandafter\def\csname LT5\endcsname{\color{black}}%
      \expandafter\def\csname LT6\endcsname{\color{black}}%
      \expandafter\def\csname LT7\endcsname{\color{black}}%
      \expandafter\def\csname LT8\endcsname{\color{black}}%
    \fi
  \fi
  \setlength{\unitlength}{0.0500bp}%
  \begin{picture}(5760.00,3528.00)%
    \gplgaddtomacro\gplbacktext{%
      \csname LTb\endcsname%
      \put(1606,704){\makebox(0,0)[r]{\strut{} 1e-06}}%
      \put(1606,998){\makebox(0,0)[r]{\strut{} 1e-05}}%
      \put(1606,1293){\makebox(0,0)[r]{\strut{} 0.0001}}%
      \put(1606,1587){\makebox(0,0)[r]{\strut{} 0.001}}%
      \put(1606,1881){\makebox(0,0)[r]{\strut{} 0.01}}%
      \put(1606,2175){\makebox(0,0)[r]{\strut{} 0.1}}%
      \put(1606,2470){\makebox(0,0)[r]{\strut{} 1}}%
      \put(1606,2764){\makebox(0,0)[r]{\strut{} 10}}%
      \put(1606,3058){\makebox(0,0)[r]{\strut{} 100}}%
      \put(1738,484){\makebox(0,0){\strut{} 0.01}}%
      \put(3247,484){\makebox(0,0){\strut{} 0.1}}%
      \put(4755,484){\makebox(0,0){\strut{} 1}}%
      \put(440,1984){\rotatebox{90}{\makebox(0,0){\strut{}$F(t,t;p)$}}}%
      \put(3584,154){\makebox(0,0){\strut{}$p$}}%
    }%
    \gplgaddtomacro\gplfronttext{%
      \csname LTb\endcsname%
      \put(3831,1323){\makebox(0,0)[r]{\strut{}$\langle F\rangle_V$}}%
      \csname LTb\endcsname%
      \put(3831,1103){\makebox(0,0)[r]{\strut{}classical simulation}}%
      \csname LTb\endcsname%
      \put(3831,883){\makebox(0,0)[r]{\strut{}kink fit}}%
    }%
    \gplbacktext
    \put(0,0){\includegraphics{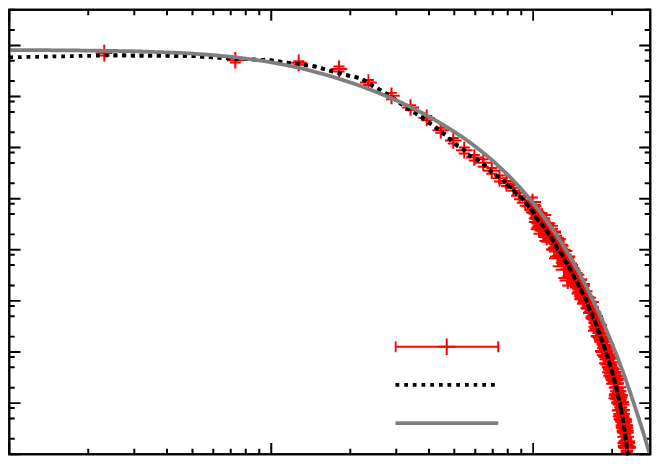}}%
    \gplfronttext
  \end{picture}%
\endgroup